\documentclass[lettersize,journal]{IEEEtran}
\usepackage{amsmath,amssymb,amsfonts}
\usepackage{graphicx}
\usepackage{textcomp}
\usepackage{xcolor}
\usepackage{filecontents}
\usepackage{booktabs} 
\usepackage{amsfonts}
\usepackage{booktabs}
\usepackage{multirow}
\usepackage{graphicx} 
\usepackage{graphics}
\usepackage{subfigure}
\usepackage{xcolor}
\usepackage{hyperref}
\usepackage{mathrsfs}
\usepackage{breqn}
\usepackage{verbatim}
\usepackage{xspace}

\usepackage{booktabs} 
\usepackage{algorithm}
\usepackage{balance}
 
\usepackage{array,multirow,graphicx}
\usepackage{float}

\newcommand{\ours}{{SHIELD}\xspace}

\newcommand{\etal}{{\em et al.}\xspace}

\newcommand{\etc}{{\em etc.}\xspace}
\newcommand{\BfPara}[1]{\vspace{0.2em}{\noindent\bf#1.}\xspace}

\usepackage{tikz}

\DeclareMathOperator*{\argmax}{argmax}

\usepackage{xcolor,colortbl}
\definecolor{darkgreen}{rgb}{0.0, 0.2, 0.13}

\newcommand{\acc}[1]{\cellcolor{black!60!white!#1}}

\newcommand{\mo}[1]{\textcolor{red}{#1}}

\definecolor{darkred}{rgb}{0.55, 0.0, 0.0}

\begin{document}

\title{SHIELD: Thwarting Code Authorship Attribution}

\author{Mohammed Abuhamad,~\IEEEmembership{Member,~IEEE,} Changhun Jung,~\IEEEmembership{Member,~IEEE,} \\David Mohaisen,~\IEEEmembership{Senior Member,~IEEE,} DaeHun Nyang,~\IEEEmembership{Member,~IEEE} \thanks{M. Abuhamad is with the Department of Computer Science at Loyola University Chicago. C. Jang and D. Nyang are with Ewha Womans University (e-mail: nyang@ewha.ac.kr). D. Mohaisen is with the Department of Computer Science at the University of Central Florida (e-mail: mohaisen@ucf.edu).}
}


\maketitle

\begin{abstract}
Authorship attribution has become increasingly accurate, posing a serious privacy risk for programmers who wish to remain anonymous. In this paper, we introduce SHIELD to examine the robustness of different code authorship attribution approaches against adversarial code examples. We define four attacks on attribution techniques, which include targeted and non-targeted attacks, and realize them using adversarial code perturbation. We experiment with a dataset of 200 programmers from the Google Code Jam competition to validate our methods targeting six state-of-the-art authorship attribution methods that adopt a variety of techniques for extracting authorship traits from source-code, including RNN, CNN, and code stylometry. Our experiments demonstrate the vulnerability of current authorship attribution methods against adversarial attacks. For the non-targeted attack, our experiments demonstrate the vulnerability of current authorship attribution methods against the attack with an attack success rate exceeds 98.5\% accompanied by a degradation of the identification confidence that exceeds 13\%. For the targeted attacks, we show the possibility of impersonating a programmer using targeted-adversarial perturbations with a success rate ranging from 66\% to 88\% for different authorship attribution techniques under several adversarial scenarios. 
\end{abstract}

\section{Introduction}\label{sec:introduction}

Code authorship attribution is the process of recognizing programmers of a given software, and there has been several works on robust and scalable attribution~\cite{abuhamad2018large,Caliskan-Islam:2015,meng2017,ALRABAEE2019S3,abuhamad2020multi}. These methods have shown that programmers can be accurately identified by their coding style, making this problem an easy task thanks to the rapid development of code analysis and machine learning techniques. Accurate attribution benefits software forensics and security, especially for identifying malicious code programmers, detecting plagiarism, and settling authorship disputes. However, the process also poses privacy risks for programmers who prefer to stay anonymous.

Recent code authorship attribution techniques heavily use machine learning models. While effective, those techniques are prone to manipulations that force the identification models to generate specific desired outputs, e.g., misclassification. One line of work for general machine learning algorithms utilized small perturbations to the input domain, resulting in adversarial examples (AEs) that are similar to the original ones, making it hard to distinguish them and posing a significant threat to machine learning models~\cite{PapernotMJFCS16}. Examining AEs in the context of authorship attributions is an under-explored topic. 


This work introduces \ours{} for generating AEs at the source code level to prevent attribution while preserving code functionality. 
Such AEs will fool a classifier into misidentifying programmers and lead to targeted attacks, e.g., imitation or mimicking. 
Investigating such capabilities by examining how prone  authorship attribution is to practical AEs allows a finer understanding of the state-of-the-art methods and helps address their shortcomings, especially with the increasing reliance on them for  identifying coding style of programmers for security applications~\cite{abuhamad2019code,Caliskan-Islam:2015,ALRABAEE2019S3}.  As a byproduct, our attacks can serve as a building block for maintaining the privacy of programmers in the presence of attribution techniques. 

\ours examines both the non-targeted and targeted attacks.
In the non-targeted attacks, known as {\em confidence reduction} or  {\em misclassification} attacks, \ours manipulates the input source code so that the identification model outputs any other author, i.e., {\em  authorship dodging} \cite{sharif2016accessorize,quiring2019misleading}, 
where \ours can use this strategy to conceal the code author identity. 
In the targeted attack, \ours manipulates the input so that the identification model outputs a specific \textit{target} author, i.e., {\em authorship imitation} or {\em evasion}, depending on the adversarial capability and objective. We apply those scenarios to various authorship attribution methods with \ours for an in-depth analysis of each method. 

Although possibly extendable to binaries~\cite{caliskan2015binary}, \ours targets source-code authorship attribution for its prevalence. 
We note that source code-level attacks are challenging, since the generated AEs should be syntactically correct, should preserve the code functionality, and should not be easily detected. 
Although those attacks can be conducted using code transformation \cite{matyukhina2019adversarial,quiring2019misleading,liu2021practical}, a process similar to author obfuscation whereby the authorship traits are hidden in the transformed code, code transformation techniques require analyzing and changing the code to the target features across various categories, including layout, lexical, syntactic, control-flow, and data-flow features.
However, a code perturbation approach provides an effective alternative for targeted and non-targeted attacks, by generating AEs to meet a specific goal without targeting the features of different categories.  

Conventionally, adversarial perturbations are applied directly to the input source and not to the feature space, e.g., perturbations in an image, not in the features extracted from that image. However, the code authorship attribution techniques are typically based on the authorship traits extracted from code, assuming a closed system where the feature extraction is not manipulated and requires perturbations at the code level. 
Therefore, attribution attacks should be designed explicitly using perturbations applied directly to the code.

\ours injects carefully chosen code samples into the target source code and then obfuscates it using an off-the-shelf obfuscator. Unlike the prior work, where obfuscation is used to conceal the identity of the code author, we use obfuscation to make it difficult for the adversary to recognize or remove the injected code parts statically.

Our threat model is more restrictive than without adopting obfuscation because code attribution must also be capable of identifying obfuscated source codes. However, it is well known that the attribution models work even in the obfuscated domain~\cite{abuhamad2018large,Caliskan-Islam:2015}. We note that while our AEs are generated at the code level, the eventual effect of the injection will be perturbations in the feature space. For convenience, we use the term perturbation to refer to injection. 

\BfPara{Contributions} Our key contributions are as follows. (1) We introduce \ours{}, a simple yet effective approach for generating AEs on code authorship attribution.  The proposed approach does not change the original code but adds carefully-crafted code blocks (i.e., perturbations) to alter the authorship attributes of the code, leading to authorship dodging and imitation.
(2) We provide a comprehensive evaluation of our technique against six state-of-the-art authorship attribution methods: DL-CAIS \cite{abuhamad2018large}, WE-C-CNN, WE-S-CNN, TF-IDF-C-CNN, and TF-IDF-S-CNN \cite{abuhamad2019code}, and Code Stylometry \cite{Caliskan-Islam:2015}. Our evaluation features a large-scale analysis of code authorship robustness under various adversarial scenarios using a dataset of 200 programmers. Our approach achieved a misidentification rate exceeding 98.5\% for non-targeted attacks while significantly reducing the confidence of the model output.  We also demonstrate imitation attacks at a rate of more than 66\% for all targeted systems when using sufficient perturbations and at 88\% when the adversary has access to samples of the targeted author.
    

\BfPara{Organization} 
In  Section~\ref{sec:identification_workflow}, we provide a brief overview of authorship attribution  workflow alongside specific details of the six targeted systems.
In Section \ref{sec:approach}, we introduce \ours{} and the attack strategy adopted in this work.
In Section \ref{sec:non-targeted_attacks}, we define the non-targeted attacks and present the associated experimental results, followed by the targeted attacks and associated results in Section~\ref{sec:targeted_attacks}. In Section \ref{sec:discussion}, we discuss our findings, including the limitations and shortcomings of our approach. 
In Section \ref{sec:related_work}, we review the related work.
We conclude our work in
Section \ref{sec:conclusion}.

\section{Code Authorship Attribution} \label{sec:identification_workflow}

\begin{figure}[t]
    \begin{center}
        \includegraphics[width=0.48\textwidth]{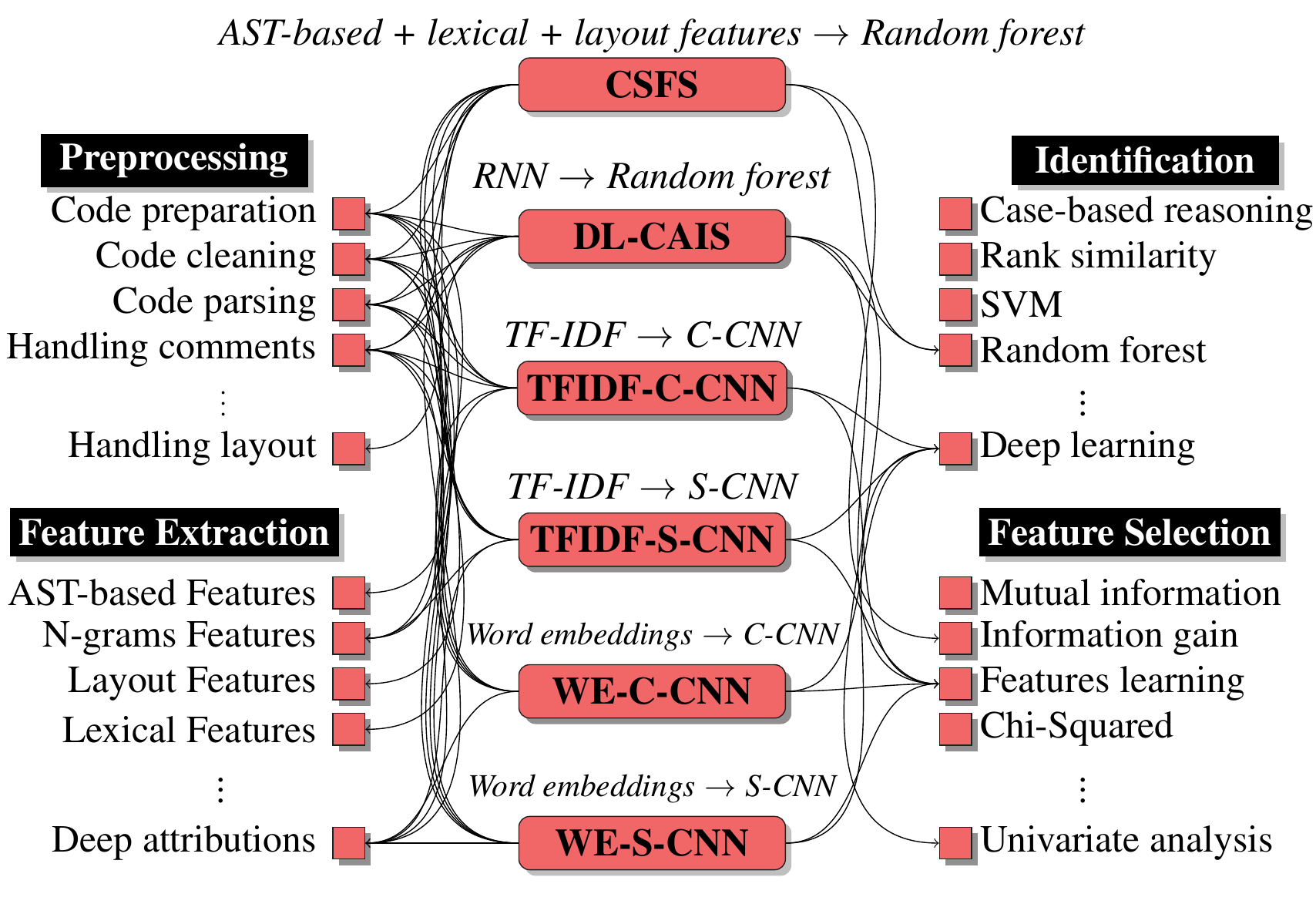}
        \vspace{-5mm}
        \caption{The workflow of code authorship attribution process, including the preprocessing, feature extraction, feature selection, and identification.}
        \label{fig:CodePrelim}
    \end{center} 
\end{figure}

\subsection{Authorship Attribution Workflow} \label{sec:preliminaries}

The workflow of the typical code authorship attribution system includes four stages: preprocessing, feature extraction, feature selection, and authorship identification. 

The {\em data preprocessing} entails removing undesired parts from the code, e.g., comments, links to external resources, \etc, and normalizing the layout features, e.g., white spaces and lines. 
The {\em feature extraction}, also known as authorship attributes extraction, requires defining distinctive coding traits. Such traits are features that capture specific characteristics of different programming styles. Typically, the features may include some or all of the following: (i) syntactic features in the program structure and implementation choices, (ii) stylometric features in variables naming, documentation, language reserved keywords usage, \etc, (iii) Layout features, e.g., whitespace characters and indentations. (iv) development environment features, e.g., platforms, editors, programming languages, \etc Feature extraction entails computing those features from the code.

The number of candidate features signify the {\em feature selection} process. The feature selection includes evaluating  authorship attributes based on, for example, the information gain or mutual information between attributes and authors to determine the prominent features. The final stage is the {\em authorship identification}, which is usually treated as a classification problem. Using the extracted (or selected) features, (supervised) classification models---such as the Support Vector Machine (SVM)~\cite{Brian2000, Burrows2014}, Random Forest Classifier (RFC)~\cite{abuhamad2018large, Caliskan-Islam:2015}, or neural networks~\cite{Burrows2014}---are trained to identify programmers. 

The workflow of the authorship attribution is shown in Figure~\ref{fig:CodePrelim}, highlighting the methods used by the six approaches investigated in this study: Code Stylometry (CSFS)~\cite{Caliskan-Islam:2015}
Deep Learning-based Code authorship identification system (DL-CAIS) ~\cite{abuhamad2018large}, TF-IDF-based concatenated CNN (TFIDF-C-CNN)~\cite{abuhamad2019code},
TF-IDF-based stacked CNN (TFIDF-S-CNN)~\cite{abuhamad2019code},
word embedding-based concatenated CNN (WE-C-CNN)~\cite{abuhamad2019code}, and word embedding-based stacked CNN (WE-S-CNN)~\cite{abuhamad2019code}.

\subsection{Our Baseline and Implementation Details} \label{implementation_details}
In the following, we review the baseline implementation of the six targeted authorship identification techniques.

\BfPara{DL-CAIS}
DL-CAIS \cite{abuhamad2018large} uses the Long Short-term Memory (LSTM) technique to learn deep authorship feature representations. The supervised feature learning model is trained to relate  the input data to the output labels, where the input in DL-CAIS is the code files initially represented as TF-IDF vectors. Namely, the input vector for a document $d_i$ to the deep learning model is represented as 
$\small
\left[\textrm{TF-IDF}(t_1, d_i, \mathcal{D}), \textrm{TF-IDF}(t_2, d_i, \mathcal{D}), \ldots, \textrm{TF-IDF}(t_n, d_i, \mathcal{D}))\right]
$,
where $n$ is the total number of terms in the corpus $\mathcal{D}$.

For dimensionality reduction, DL-CAIS uses the order of term frequency to select features.
For every term $t_i$ and document $d_i$, DL-CAIS calculates
$x_i = \bigcup \textrm{TF-IDF}(t_i, d_j, \mathcal{D}) \quad \forall~j: 1\leq j\leq \|\mathcal{D}\|$,
where $\cup$ is a feature selection operator.

Following the implementation in the original work obtained from~\cite{abuhamad2018large}, we used the top 2,500 TF-IDF features as a single step vector representation to the LSTM model for learning the deep authorship features. The LSTM architecture includes three layers of 128 LSTM units followed by three fully-connected layers with 1024 units connected to a softmax layer with the number of programmers (classes). 
In LSTM training, a model is created using input code samples and the corresponding authors (pairs) by  minimizing the softmax-cross-entropy loss.
The LSTM model is then used to generate deep authorship feature representations of code samples as the output of the second-last layer. Those features are used 
to build an RFC with 150 trees grown to the maximum extent.

\BfPara{Code Stylometry} Caliskan-Islam \etal \cite{Caliskan-Islam:2015} introduced Code Stylometry Feature Set (CSFS) for authorship attribution using three sets: lexical and layout features extracted directly from the source code, and syntactic features extracted from the Abstract Syntax Tree (AST) of the source code. 
The lexical features express the programmer's preferences for using functions, nesting depth, and specific expressions and keywords, while the layout features are statistics about whitespace characters and indentation.
The syntactic features represent properties of the program structure, e.g., the depth and frequency of an AST node, node unigrams and bigrams frequency, \etc
In total, CSFS generated +120k features for 250 authors with nine files each. This requires a feature selection process, which is done using information gain by scoring features based on the difference in the Shannon entropy of classes' distribution and the conditional distribution of classes given a feature. The feature score is 
$\text{IG}(y, c_i) = E(Y) -E(y|c_i)$, where $y$ is a class label, $E$ is the Shannon entropy, and $c_i$ is the $i\text{-th}$ feature. 

The resulting features with non-zero gain are denoted by IG-CSFS. The top-{\em k} features are used.
Using the implementation of Caliskan-Islam \etal \cite{Caliskan-Islam:2015}, the input code samples are represented with the top 1024 IG-CSFS authorship features to build RFC with 150 trees grown to the maximum extent.

\BfPara{CNN-based Systems}
Abuhamad \etal~\cite{abuhamad2019code} used  word embedding and TF-IDF as input methods for CNN-based models with various architectures.
For word embedding, a code sample $x_i \in \mathbb{R}^{1\times (n\times d)}$ is represented as a sequence of expressions, $x_i^{t_1:t_n} = t_1 \oplus t_2 \oplus \dots \oplus t_n$, where $t_i \in \mathbb{R}^{d}$ is the representation of the $i\text{-th}$ term in $x_i$ and $\oplus$ is the concatenation operator. 
For efficiency, they used a predefined sequence length $n$ for code samples where short sequences are padded, and long sequences are truncated.  
For the TF-IDF representation, a similar approach was adopted as in DL-CAIS~\cite{abuhamad2018large}.

For identification, 
Abuhamad \etal~\cite{abuhamad2019code} used two CNN architectures, concatenated and stacked. 
The concatenated CNN includes concatenating the feature maps of a number of convolutional layers, while the stacked CNN follows the typical CNN model by applying convolutional layers on top of each other to generate feature maps based on features of the previous layer. 
For both architectures, a softmax classifier is used---the softmax function, defined as $\mathsf{softmax}(y_i)= \frac{e^{y_i}}{\sum_j e^{y_j} \forall~ j}$, signifies the probability of classifying the class at the $i$-th position among all class labels---and the output is the class with the highest probability. The following are the implementation details of four targeted CNN-based systems.

\begin{itemize}
\item {\bf WE-C-CNN:} We set the word embedding matrix to generate 128-word representations. 
The input code samples are then represented as a matrix of $n \times 128$ representations, where $n$ is the number of terms in a code sample. 
We set $n=256$ to fix the length representations to the convolutional layers as short samples are padded with zeros, and long samples are truncated.
For the convolutional layers, three layers of 128 filters of different sizes (3, 4, and 5) are adopted to receive the input representation. 
The outputs of the three layers are concatenated and connected to a softmax layer with the same size as the number of the considered classes.

\item {\bf WE-S-CNN:} Similar to the settings of WE-C-CNN, the code samples are represented as $x \in \mathbb{R}^{256 \times 128}$ since the word embeddings are set to be 128 vector representations. 
The CNN model architecture follows the typical stacked CNN layers with three consecutive layers, each with 128 filters of different sizes (3, 4, and 5). 
For this architecture, we used max-pooling after each layer to reduce the size of feature maps produced by the convolutional layer. 
The last max-pooling layer is connected to the output softmax layer.

\item {\bf TFIDF-C-CNN:} The input code samples are represented with the top 2500 TF-IDF features and fed to 1-dimensional convolutional layers. 
Similar to WE-C-CNN, we use three filter sizes (3, 4, and 5) and 128 filters per layer. 
The feature maps produced by the three convolutional layers are then concatenated and connected to a softmax layer with the same size as that of the number of classes.

\item {\bf TFIDF-S-CNN:} Similar to TFIDF-C-CNN, the input code samples are represented with the top-2500 TF-IDF features and fed to three stacked 1-dimensional convolutional layers. 
All layers consist of 128 filters and each layer has a different filter size.  We used max-pooling for the stacked CNN architecture and the last max-pooling layer is connected to a softmax layer with the same size as that of the number of classes.
\end{itemize}

\BfPara{Deep Learning Training}
Both RNN-based and CNN-based models are trained using the Adam optimizer with a static learning rate of $10^{-4}$ to minimize the softmax-cross-entropy loss, since all models are trained in a supervised manner. 
In all cases, the training process is terminated after 1,000 iterations. 
To prevent overfitting, we used the {\em dropout regularization} with keep-rate of 70\% and {\em L2 regularization} with $\lambda$ regularization strength of $10^{-3}$. 
Moreover, we used a mini-batch approach with a mini-batch size of 64 observations in the training process of all deep learning-based architectures.

\begin{figure*}[tb]
\centering
\begin{subfigure}[Original source code \label{fig:Original_code}]{\includegraphics[width=0.24\textwidth]{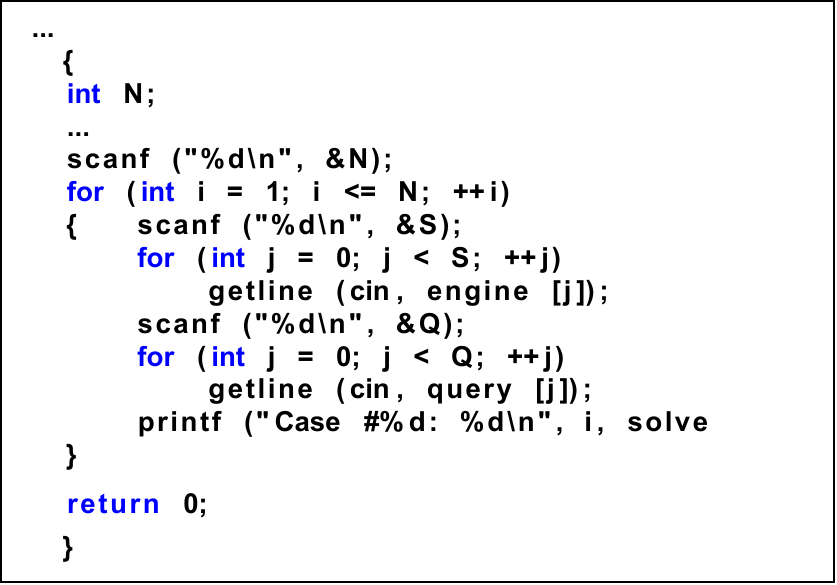}} 
\end{subfigure}
\begin{subfigure}[Adversarial example (1) \label{fig:AE1}]{\includegraphics[width=0.24\textwidth]{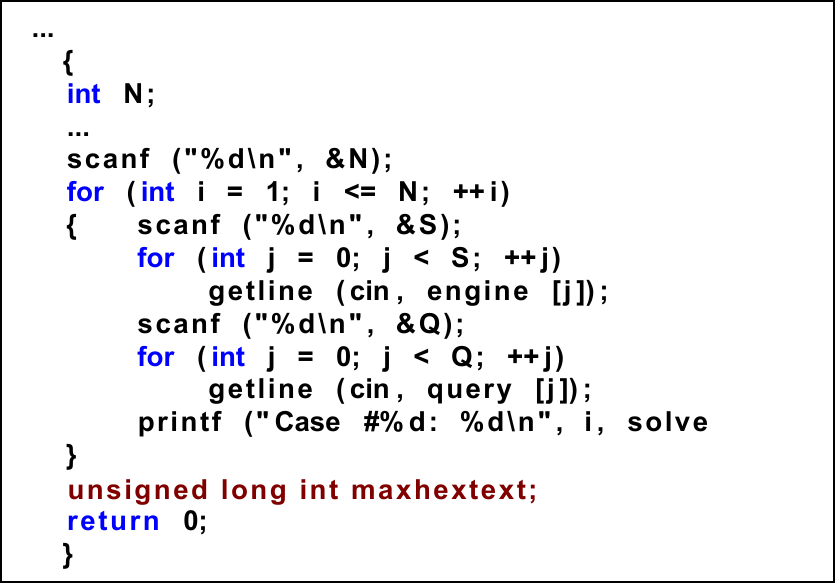}} 
\end{subfigure}
\begin{subfigure}[Adversarial example (2) \label{fig:AE2}]{\includegraphics[width=0.24\textwidth]{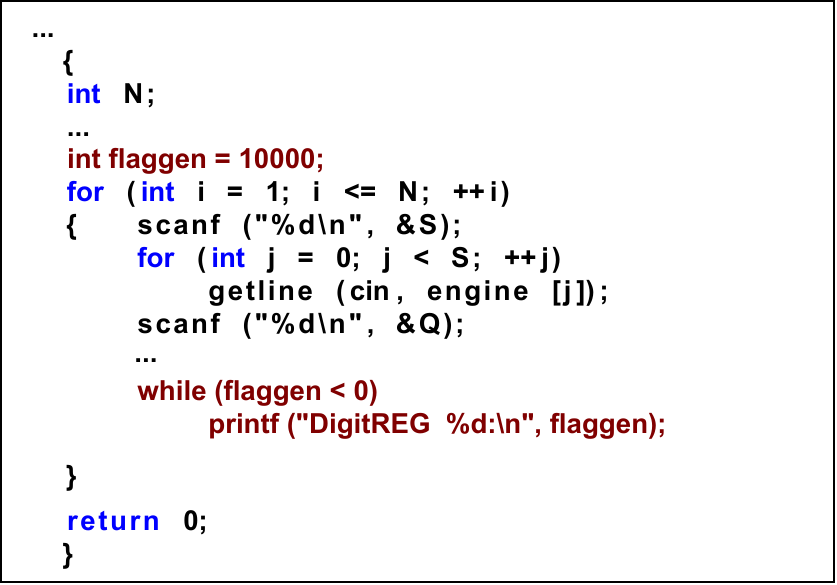}} 
\end{subfigure}
\begin{subfigure}[Adversarial example (3) \label{fig:AE3}]{\includegraphics[width=0.24\textwidth]{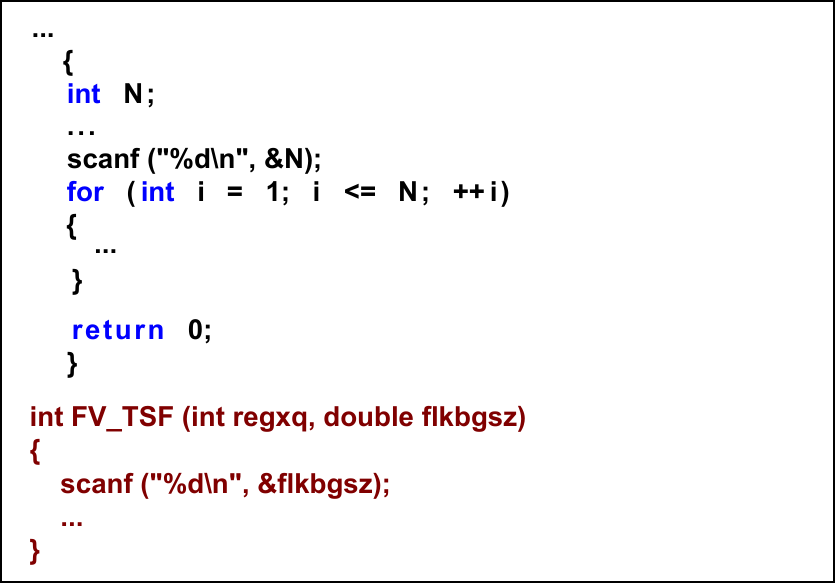}} 
\end{subfigure}\vspace{-3mm}
\caption{Example of adversarial source-code examples. The code perturbations can be variable declarations (Adversarial example 1), loops and control statements (Adversarial example 2), functions (Adversarial example 3). Note that the added perturbations (in \textcolor{darkred}{red color}) are not meant for execution.}
\label{fig:examples} \vspace{-4mm}
\end{figure*}

\section{\ours{}: Methods} \label{sec:approach} 

This section describes our proposed attack strategy and the methods employed for the adversarial code perturbation.

\subsection{The Adversarial Attack Strategy}
Our robustness assessment of the authorship attribution systems includes exploring the characteristics of their behavior under adversarial settings. 
\ours{} introduces code perturbations to implement four threat models categorized by their target specificity and adversarial knowledge. 
All of the attacks are implemented under the {\em black-box} threat model.

The adversary starts by defining the source of perturbation, i.e., code statements, expressions, variable names, \etc, to be used for generating the adversarial example. Based on the threat model, selecting the source of perturbations can be through random selection or targeted selection. 
This study investigates four threat models, namely the non-targeted attacks (Section~\ref{sec:non-targeted_attacks}) and the targeted attacks T1 through T3 (Section~\ref{sec:targeted_attacks}). The non-targeted attacks and targeted attacks T1 use code perturbations that are generated through a random selection of code statements and expressions, while the targeted attacks T2 and T3 use targeted selection of code perturbation.
After defining the source of perturbation, the code perturbations are generated from a template library that includes code statements and blocks that follow the syntax rules of the programming language (e.g., C++ in our experiments). 

We designed a code template library with a list of operational and control statements. Using the perturbation expressions, the code statements produced by the templates are injected directly into the original code or used as basic units for a larger code block, e.g., functions/methods. More details about the code perturbation are in Section \ref{CodePerturbation}.

After adding the code perturbation to the code sample to defeat identification, \ours{} aims to hide the added perturbation using {\bf code-to-code obfuscation}~\cite{stunnix,tigress}. Assuming working with powerful authorship identification models that can identify and detect authorship traits despite obfuscation, \ours{}  incorporates obfuscation to hinder the detectability of perturbation and easy-to-perform static analysis. Previous studies \cite{abuhamad2018large,Caliskan-Islam:2015} have shown that identifying authors of obfuscated code can be accurately achieved.

The AE is fed into the code authorship system for feature extraction and identification. Since \ours{} assumes a {\em black-box} model, it optimizes the perturbation by repeatedly querying the system and observes the changes in the output distribution. Using the identification model's probability distribution, \ours{} optimizes the perturbation through the changes in the code expressions and statements used to generate the attacks on the attribution system. 

The number of queries that \ours{} conducts in order to succeed depends on the threat model, \ours{}'s specific objectives, and level of knowledge. 
We describe the threat models in detail in Section \ref{sec:threatModel1} (non-targeted attacks) and Section \ref{sec:threatModel2} (targeted attacks).
We observed that launching successful non-targeted attacks, even with small random perturbations, requires fewer rounds of queries with the identification model compared to the targeted attacks. 

For evaluation, we defined three threat models for the {\em targeted attacks.} {T1:} This model requires a random selection of perturbation for targeted attacks. {T2:} This model requires accessing two samples of the target programmer that are not included in the training dataset used for the baseline model. {T3:}  This model extends the knowledge of the adversary to code samples from a suspect set of programmers. The adversarial objective of T3 is to generate code samples that imitate the style of the closest programmer.

The adversary gains more capabilities by accessing samples from the target set using T2 and T3. Such capabilities include observing the most representative features of the target and enhance the code perturbation to meet an adversarial goal. 
For T2,  the adversary applies a targeted selection of perturbations (i.e., code statements, variable names, \etc) based on the target's samples. The adversary applies the targeted selection over samples from multiple programmers to launch T3.

\begin{figure}[t]
\centering
\includegraphics[width=0.3\textwidth]{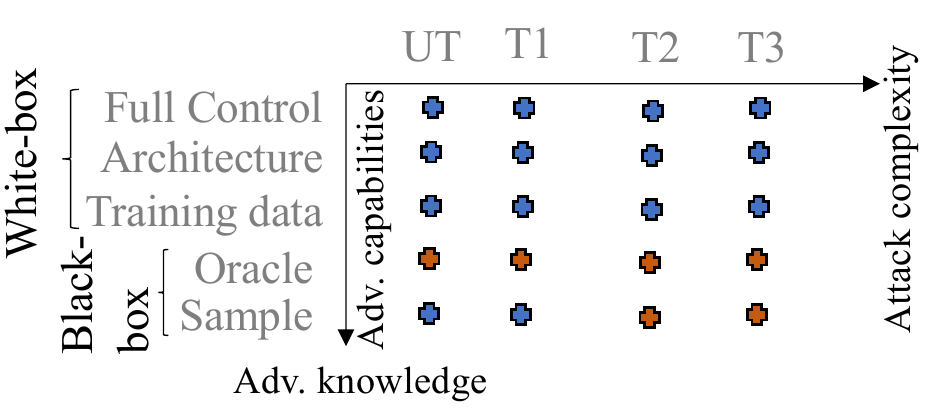}\vspace{-4mm}
\caption{{A general categorization of the attacks from the perspective of the adversary's capabilities and goals. This study explores the orange-colored attacks as they are more relevant to the addressed problem space (UT stands for untargeted, while T1 through T3 are targeted attacks). }}
\label{figure:AML_Preliminary}\vspace{-3mm}
\end{figure}

Figure \ref{figure:AML_Preliminary} shows the adversarial capabilities and objectives of the different threat models explored by \ours{} categorized by the attack specificity (targeted and non-targeted attacks). 
Note that all scenarios are under the {\em black-box} assumption.
Therefore, the adversary's capabilities include only {\em oracle} and/or {\em samples}. For the oracle access, the adversary has no knowledge about the model, but the ability to access an oracle allows her to conduct queries to the model and infer the relation between inputs and outputs. For sample access, the adversary has access to several samples from the victim that can be used to enhance the adversarial perturbations. 

\subsection{Adversarial Code Perturbation} \label{CodePerturbation}
Introducing perturbations on the feature representations assumes a \textit{white-box attack} where the adversary has full control and access to all stages of the system operation, limiting the approach's practicality. 
Assuming that the adversary has limited access to the system (i.e.,  {\em black-box attack}), we investigate  perturbations to the input code directly.

\BfPara{Code Perturbations}
Code perturbations can be applied as variable declarations, loops, and control statements, functions, \etc Figure~\ref{fig:examples} shows adversarial examples.
For an author that tries to evade identification, this approach is implemented by the author with full control over the source code. 
For code functionality preserving, the added perturbations are not meant for execution, and we achieve that by adding code perturbations as methods (functions) that never get called or executed in the main code, or as negative conditional statements.  To enhance the perturbations, we: (1) limit the size of perturbation to the minimal size that enables the attack, (2) ensure the perturbation code inherits the flow of the syntax rules of the language, (3) ensure the statements of the adversarial code should maintain the same size as other statements in the original code, and (4) ensure that the variable names should follow the naming rules and convention of the language. 

\BfPara{Targeted Programming Language} Our dataset includes code samples written in C++, and the code perturbations are presented as C++ functions that follow the C++ function structure with a function body that includes multiple statements.  
The six major statement types that we considered in our implementation are as follows. 
(1) Variable declaration and assignment. 
(2) Arithmetic, relational, and logical operations. 
(3) IF and IF-ELSE statements. 
(4) SWITCH statement. 
(5) FOR loop statement. 
(6) DO-WHILE, WHILE loop statement. 

Each statement includes at least one line of code, e.g., a declaration, an assignment, or an operation. 
The selection of statement types and the naming criteria for variables and functions in the code perturbation are based on the underlying assumptions of the attack. 
Assuming a black-box scenario for non-targeted and targeted attacks, variable names are generated randomly with a predefined variable-naming template. 
However, assuming more adversarial knowledge in the imitation attack, the considered names can be selected from previously observed used names by the targeted programmer.
We note that generating random names might cause the out-of-vocabulary problem for approaches that use TF-IDF or word embeddings for representing the code sample.

\section{Non-targeted Attack}  \label{sec:non-targeted_attacks}
This section describes the non-targeted attacks and shows their impact on the performance of authorship attribution systems. We describe the threat model, explore the baseline performance of the six targeted attribution systems, and then evaluate the attack impact on their performance.

\subsection{Threat Model 1: Non-targeted Attack}\label{sec:threatModel1}
For a non-targeted attack, we assume a \textit{black-box} scenario for accessing the code authorship attribution system. The adversary sends a source-code file to the attribution system and receives the predicted output and the confidence score. The adversary does not have access or knowledge about the components of the system, including the training data, feature extraction/selection methods, and the model structure and design. 
The adversarial objective of a non-targeted attack is to mislead the authorship identification model to predict a wrong programmer. The adversary tries to reduce the confidence of the model through the manipulation of authorship attributes using code perturbation. This kind of attack is referred to as a \textit{confidence reduction attack} and \textit{dodging attack}.

\BfPara{Model Definition}
The confidence of an identification model is defined as the probability of predicting a programmer given a test sample.
Since we are targeting various systems with different classification techniques, the
confidence of the targeted approaches is defined as follows. 
DL-CAIS and CSFS adopt RFC where the confidence is the number of decision trees voting for a given class. 
For an author $y_i \in \mathcal{Y}$, the identification confidence is the percentage of trees voted for $y_i$ to be the programmer who wrote a given code sample $x_i \in \mathcal{X}$, where $\mathcal{Y}: \{y_0,y_1, \dots, y_m\}$ is the suspect set and $\mathcal{X}: \{x_0,x_1, \dots, x_n\}$ is the collection of samples from the test set. Therefore, the confidence is estimated as: 
$$ \textit{conf~}(y_i) = \sum_{j}{\textit{vote}_j(y_i)} \times {\|T\|}^{-1},$$ 
where $\textit{vote}_j(y_i)$ is the $j\text{-th}$ tree voting for $y_i$ and $\|T\|$ is the total number of trees in RFC. 

On the other hand, CNN-based models (WE-C-CNN, WE-S-CNN, TFIDF-C-CNN, and TFIDF-S-CNN) adopt softmax classifier where the confidence is the softmax score for a given class. 
For an author at the $i$-th position of $y$, the identification confidence is estimated as: $\textit{conf~}(y_i)= \text{softmax}(y_i)$.

\BfPara{The Adversarial Objective} This attack aims to delude the identification model by manipulating the input code files so that authorship attributes become ambiguous. More precisely, the goal is to decrease the confidence in the models' predictions to lead the model to misidentification or prediction rejection.  
This is done by adding code perturbation $\delta$ to the input $x$, such that the generated adversarial  
code $\bar{x} = x + \delta$ serves the purpose of minimizing the confidence with at least $\varepsilon > 0$, {where $\varepsilon$ the confidence reduction level}. 
We note that $\delta$ is the code perturbation on the source-code input level. 

The objective is to find a minimum $\delta$ disturbing more $\varepsilon$: 
\[\small
f_{\delta, \varepsilon} (x) = \min_{\delta} \{|\delta| \quad s.t. \quad \{ \textit{conf~}(y|x) -  \textit{conf~}(y|\bar{x})\} \geq \varepsilon \}.
\] 
The $\textit{conf~}(y|x)$ indicates the probability of correctly assigning $x$ to the rightful programmer $y$.  
If $ \textit{conf~}(y|\bar{x}) < \textit{conf~}(\bar{y}|\bar{x})$, then the AE $\bar{x}$ is attributed to $\bar{y}$ (misidentification).


\begin{figure}[tb]
\centering
\begin{subfigure}[50 programmers \label{fig:Misidentification50}]{\includegraphics[width=0.14\textwidth]{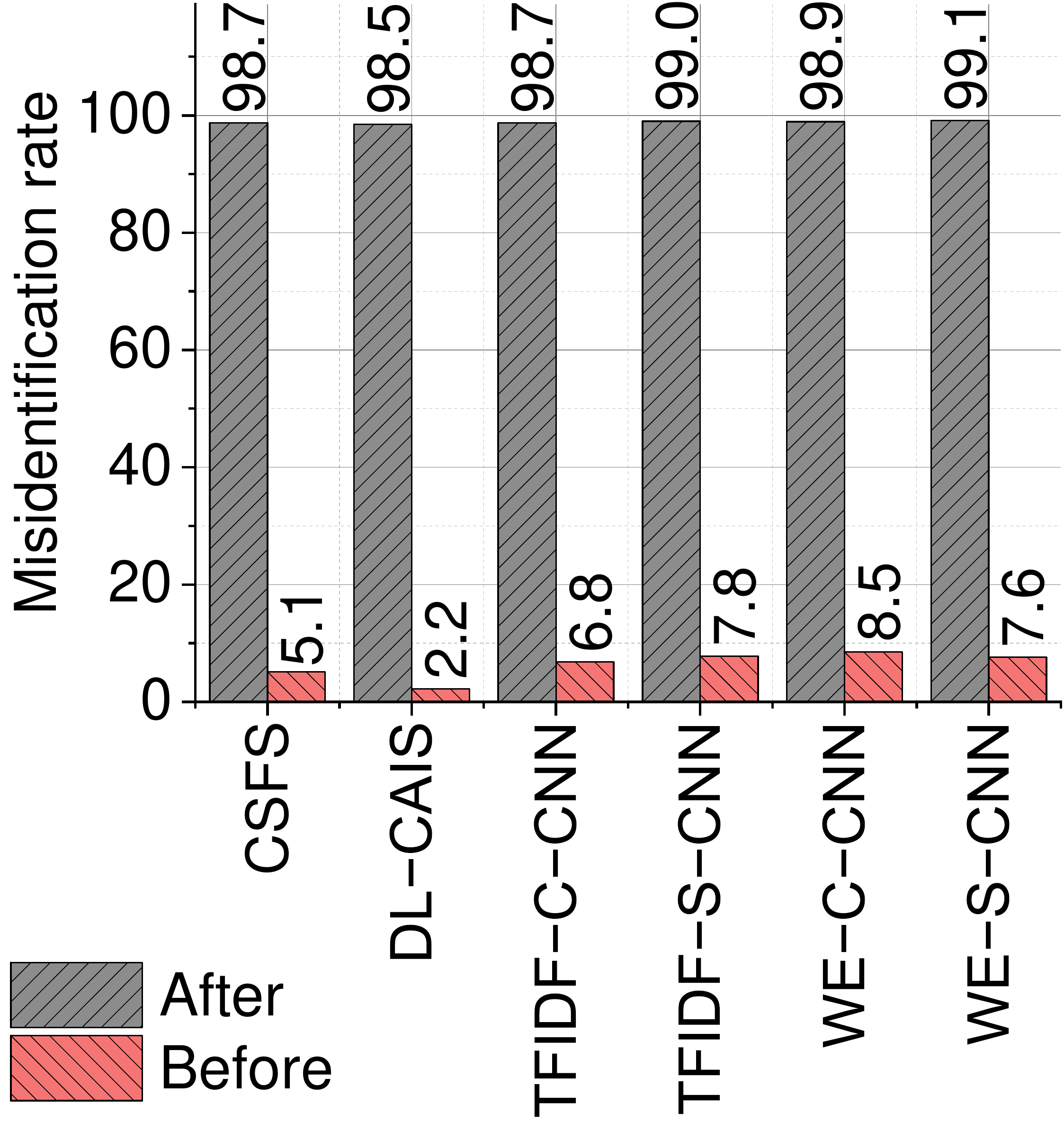}} 
\end{subfigure}
\begin{subfigure}[100 programmers \label{fig:Misidentification100}]{\includegraphics[width=0.14\textwidth]{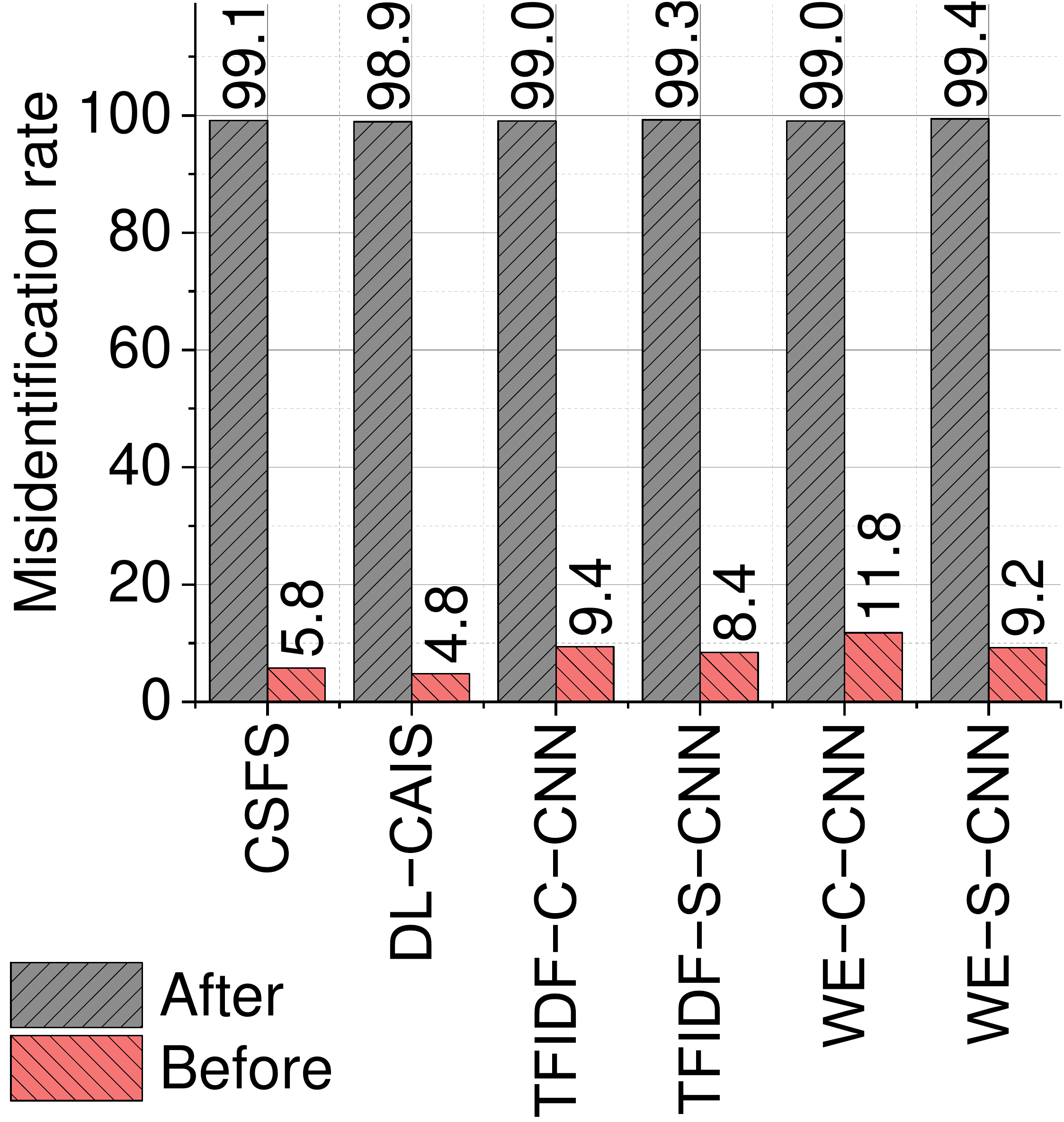}} 
\end{subfigure}
\begin{subfigure}[200 programmers \label{fig:Misidentification200}]{\includegraphics[width=0.14\textwidth]{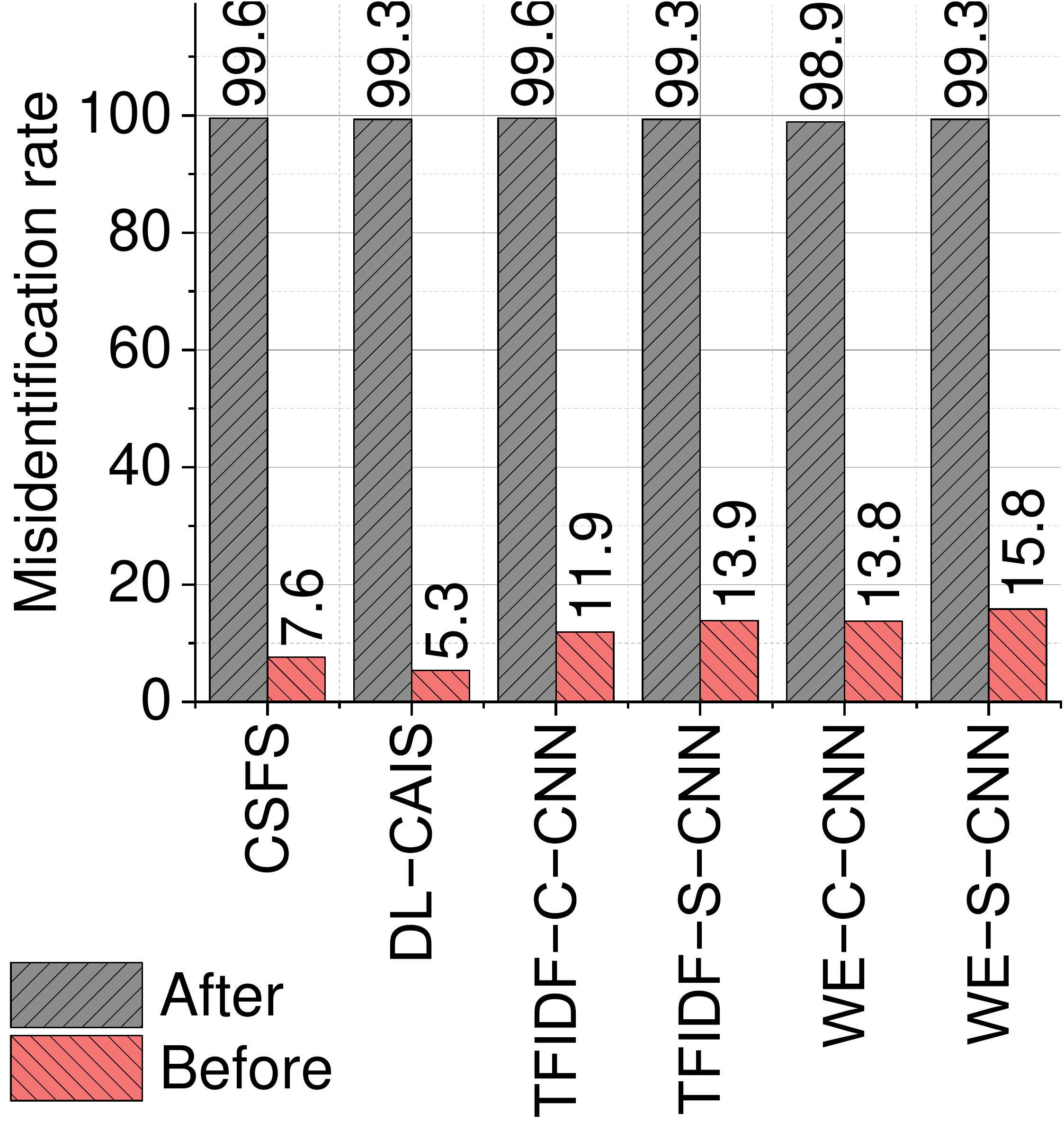}} 
\end{subfigure}
\caption{{The misidentification rate of targeted systems before and after the attacks using different datasets. The baseline is obtained using 9-fold cross-validation.}}
\label{fig:untargeted_attack}
\end{figure}

\begin{figure}[tb]
\centering
\begin{subfigure}[50 programmers \label{fig:Confidence50}]{\includegraphics[width=0.14\textwidth]{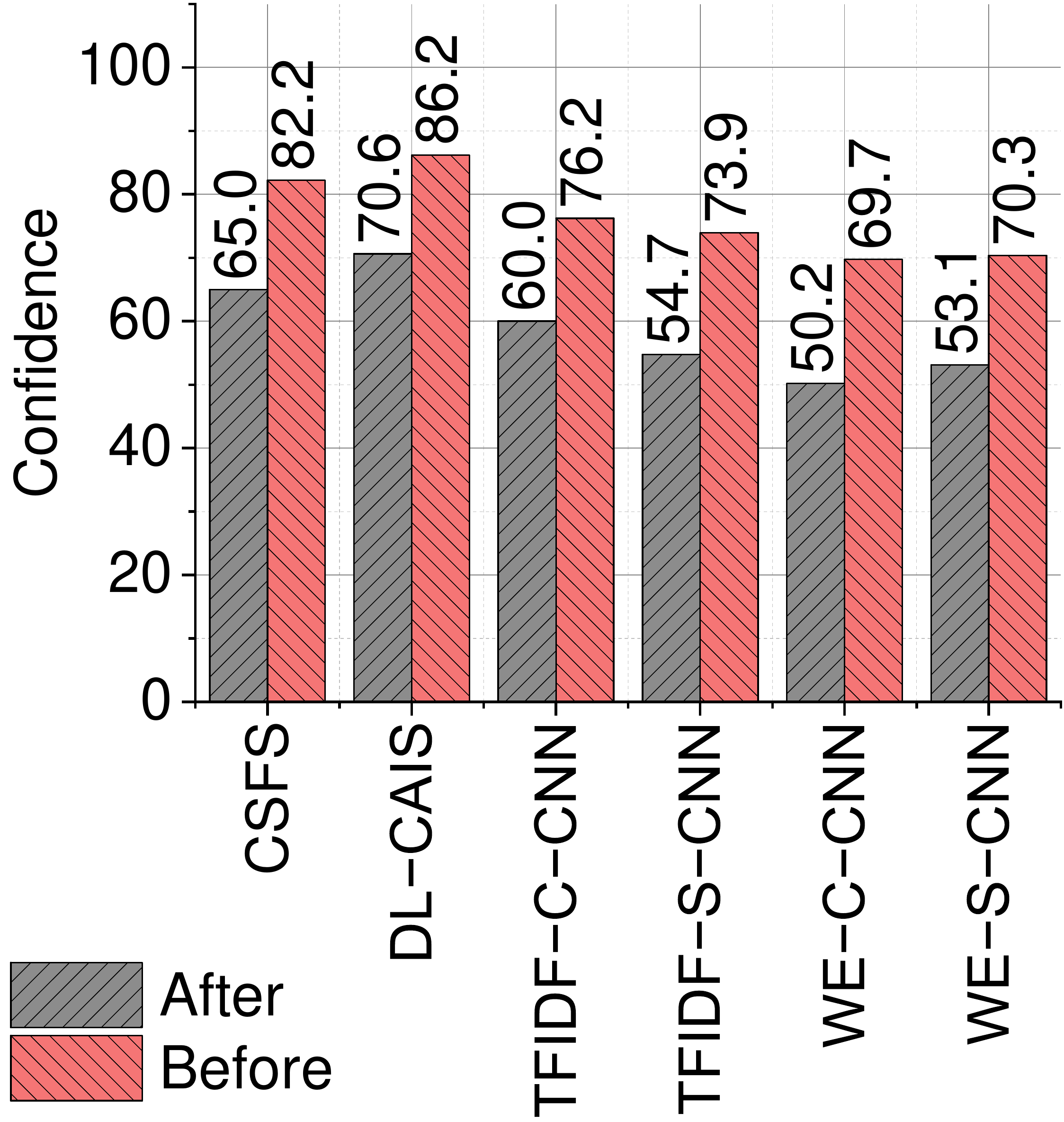}} 
\end{subfigure}
\begin{subfigure}[100 programmers \label{fig:Confidence100}]{\includegraphics[width=0.14\textwidth]{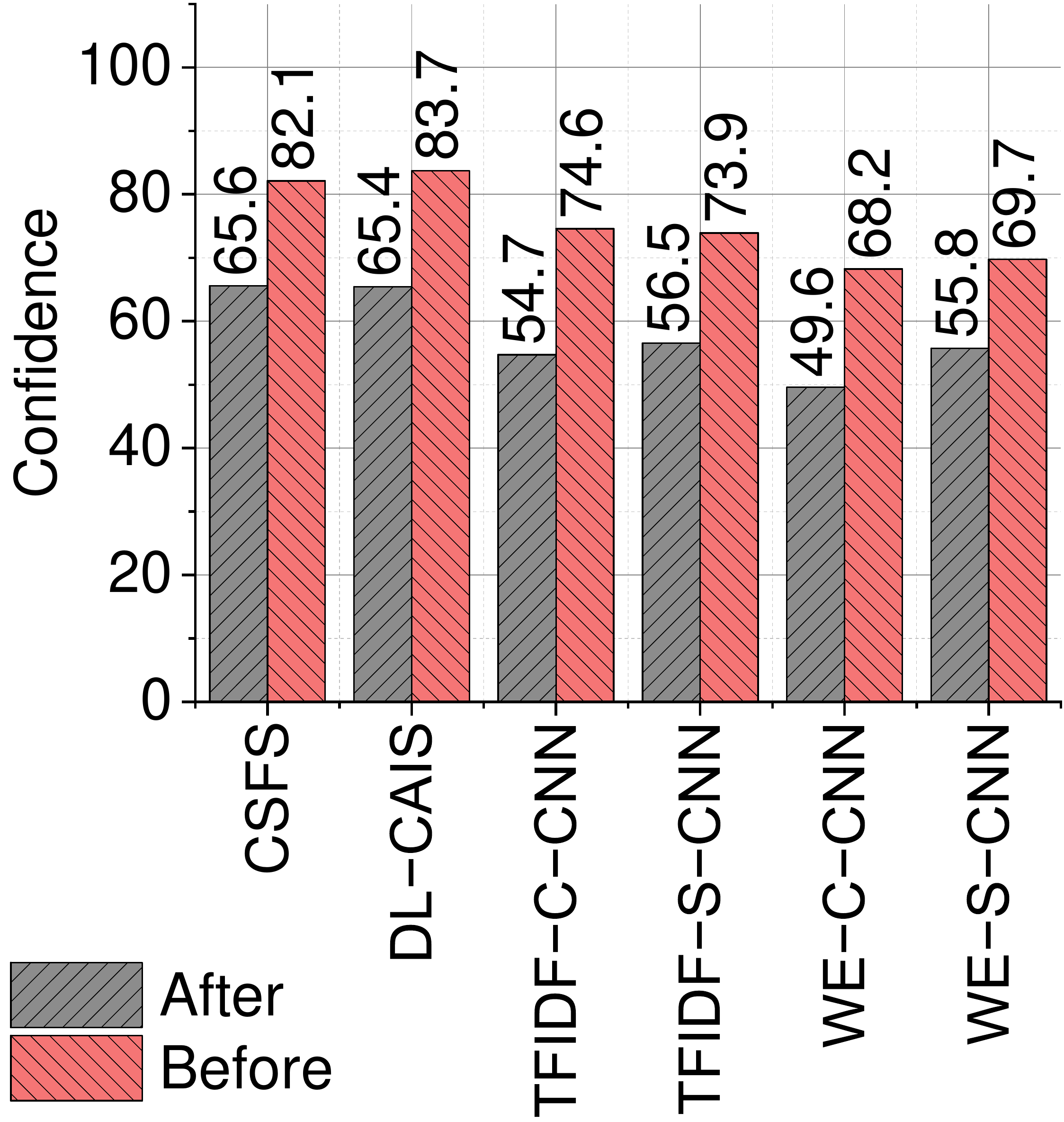}} 
\end{subfigure}
\begin{subfigure}[200 programmers \label{fig:Confidence200}]{\includegraphics[width=0.14\textwidth]{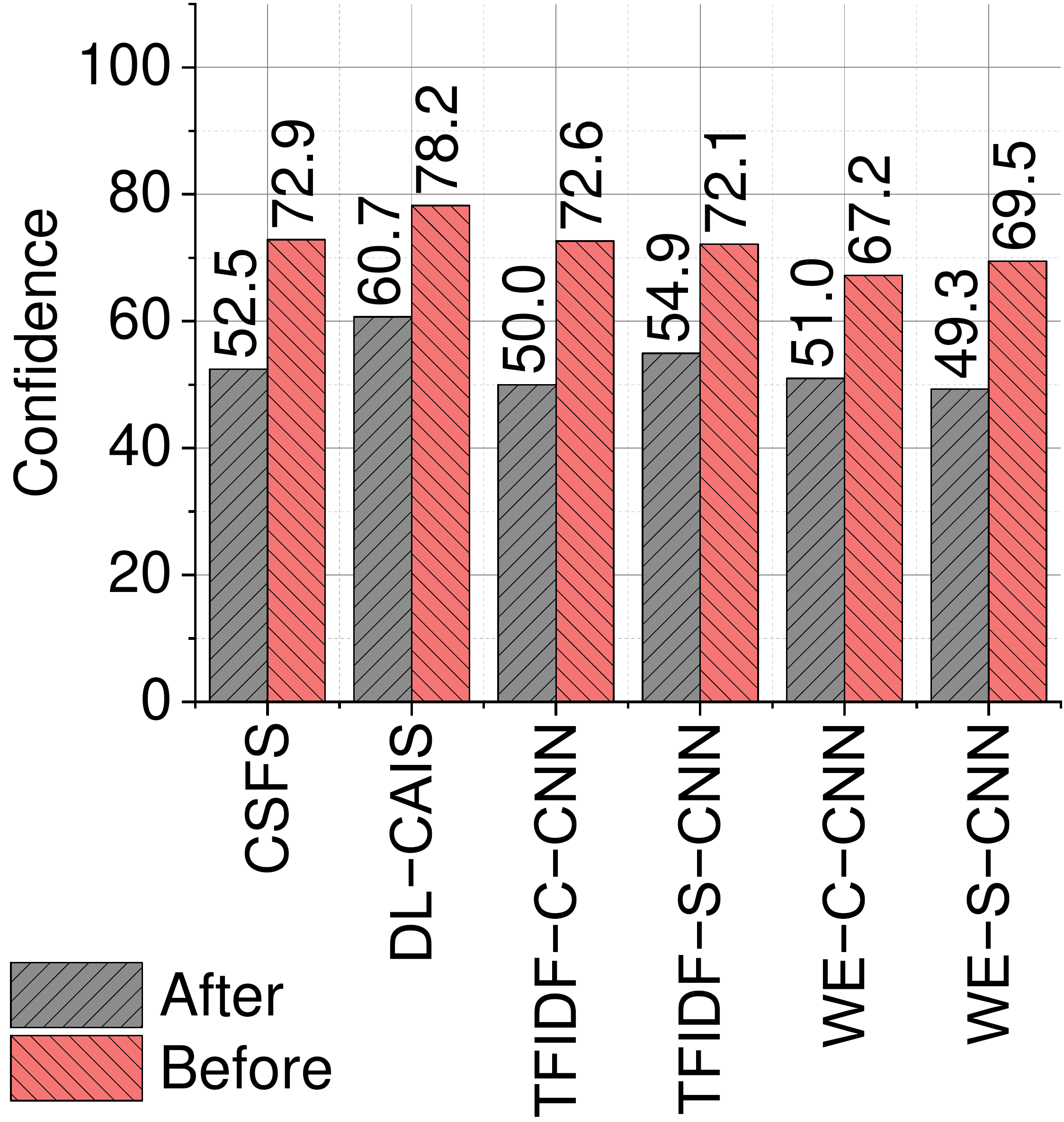}} 
\end{subfigure}\vspace{-3mm}
\caption{{The confidence of models for the predicted class before and after the attacks. The size of the dataset impacts the confidence of the models.}}
\label{fig:ConfidenceReduction} \vspace{-4mm}
\end{figure}


\BfPara{Adversarial Capabilities} 
For achieving the adversarial goal of non-targeted attacks under the \textit{black-box} scenario, 
the perturbation $\delta$ is generated randomly regardless of the code representation scheme adopted by the targeted system. The adversary only sends the adversarial example and receives the prediction output and score. The adversary iteratively enhances the adversarial perturbation until achieving the adversarial objective.    
We note that the adversary gains more advantage when accessing the training data since knowing the space of code expressions used in the dataset and their impact on the authorship attribution reduces the size of $\delta$ significantly. However, we only assume a \textit{black-box} scenario for this attack.

\subsection{Experiments: Baseline Evaluation} \label{sec:baseline_evaluation_nontargeted}

\BfPara{Our Dataset}
The evaluation is conducted using Google Code Jam (GCJ) competition~\cite{CodeJam}. 
GCJ is an international programming competition run by Google since 2008. At GCJ, programmers use several programming languages and development environments to solve programming problems over multiple rounds.
The most common languages at GCJ are C++, Java, Python, and C, in order. For this work, we use a dataset of 200 C++ programmers with nine code samples \textcolor{black}{that have {68.42} lines of code on average}. The dataset of C++ samples is collected from GCJ competition across all years from 2008 to 2016. The number of code samples, programmers, and years are consistent with prior work.

\BfPara{The Evaluation Metrics}
We evaluate the success of non-targeted attacks based on the degradation in the model confidence and the misidentification rate.
When the predicted label $y_k$, i.e.,  at the $k$-th index where $\argmax_k P(\mathcal{Y}|\bar{x}_i)$ occurs, for the adversarial code sample 
$\bar{x}_i$ is not the same as the correct class label $y_i$ of the original code sample $x_i$, this results in misidentification.
The misidentification rate is simply calculated as: 
$\frac{1}{n}\sum_i^{n} I(y_k \neq y_i),$
where $I$ is the count operator, $y_k$ is the predicted label, and $y_i$ is the actual label.



\BfPara{Baseline Identification Results}
Figure~\ref{fig:untargeted_attack} shows the identification accuracy achieved by the targeted systems using different sub-datasets with different numbers of programmers using 9-fold cross-validation evaluation.  
The results show that 
RNN-based deep representations adopted by DL-CAIS have enabled an identification accuracy ranging from 94.65\% for identifying 200 programmers to 97.8\% for identifying 50 programmers. 

Code stylometry features adopted in CSFS 
have achieved an identification accuracy between 92.4\% and 94.9\% for different datasets.
Using word embeddings to represent code files for the CNN-based approaches, WE-C-CNN and WE-S-CNN have achieved an identification accuracy of 86.23\% and 84.18\% for identifying 200 programmers, respectively.
TF-IDF representations with CNN enhanced the identification accuracy, reaching 88.1\% and 86.12\% for identifying 200 programmers using 
TFIDF-C-CNN and TFIDF-S-CNN, respectively.

Figure~\ref{fig:ConfidenceReduction} shows a high identification accuracy is achieved by most systems accompanied by a high identification confidence, where the confidence exceeds 67\% in all settings for all identification techniques.
For systems that adopt RFC, such as DL-CAIS and CSFS, the model confidence is higher than in systems that utilize a softmax layer for classification, such as the CNN-based systems (e.g., WE-C-CNN). This is because the confidence of RFC models is defined by the number of trees in the random forest that vote for a given class label, while the softmax layer captures the probability distribution of assigning the input data to all classes. In our experiments, the identification confidence decreases as the number of programmers increases, as shown in Figure~\ref{fig:ConfidenceReduction}.

\begin{figure*}[tb]
\centering
\begin{subfigure}[CSFS \label{fig:untargeted_CSFS}]{\includegraphics[width=0.155\textwidth]{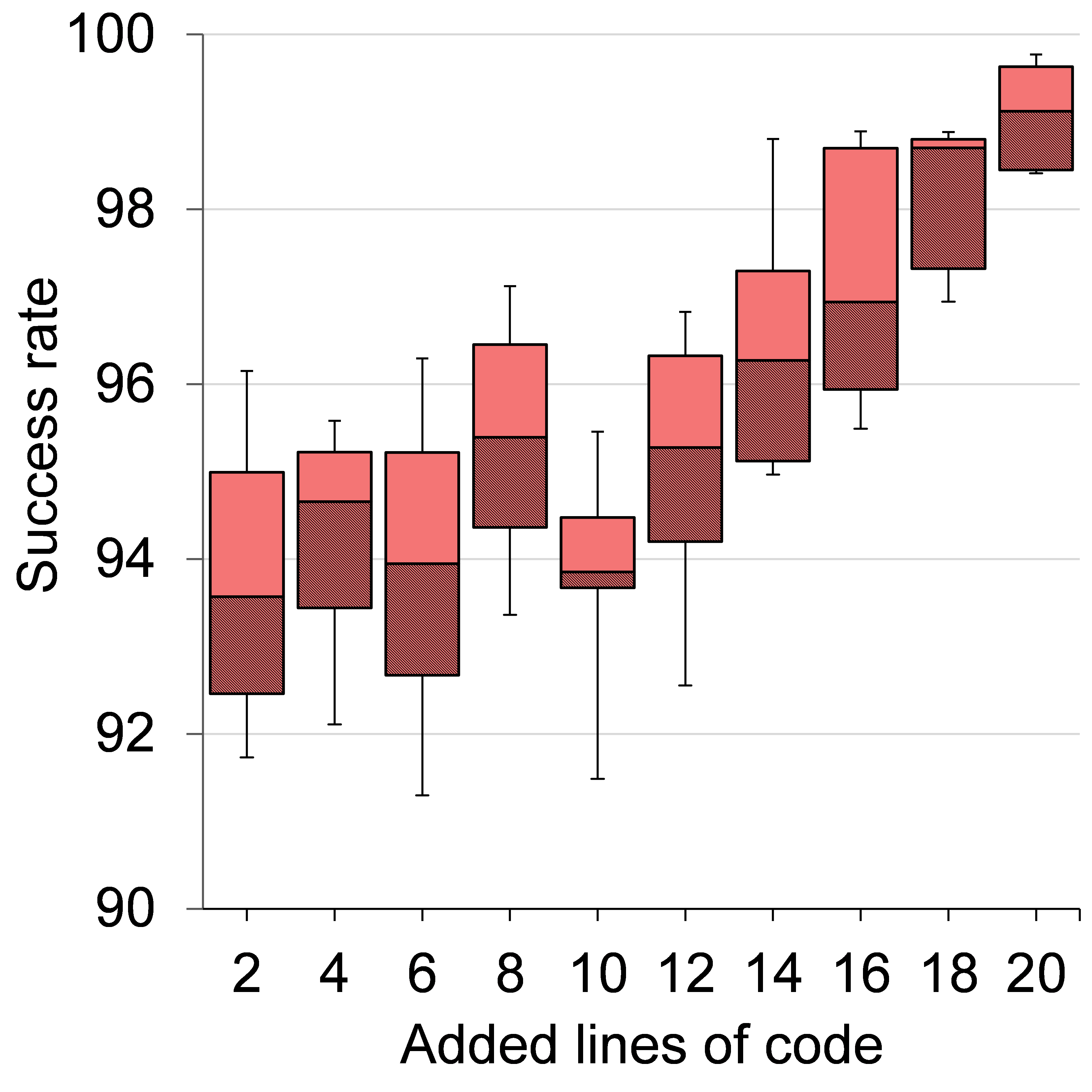}} 
\end{subfigure}
\begin{subfigure}[DL-CAIS \label{fig:untargeted_DLCAIS}]{\includegraphics[width=0.155\textwidth]{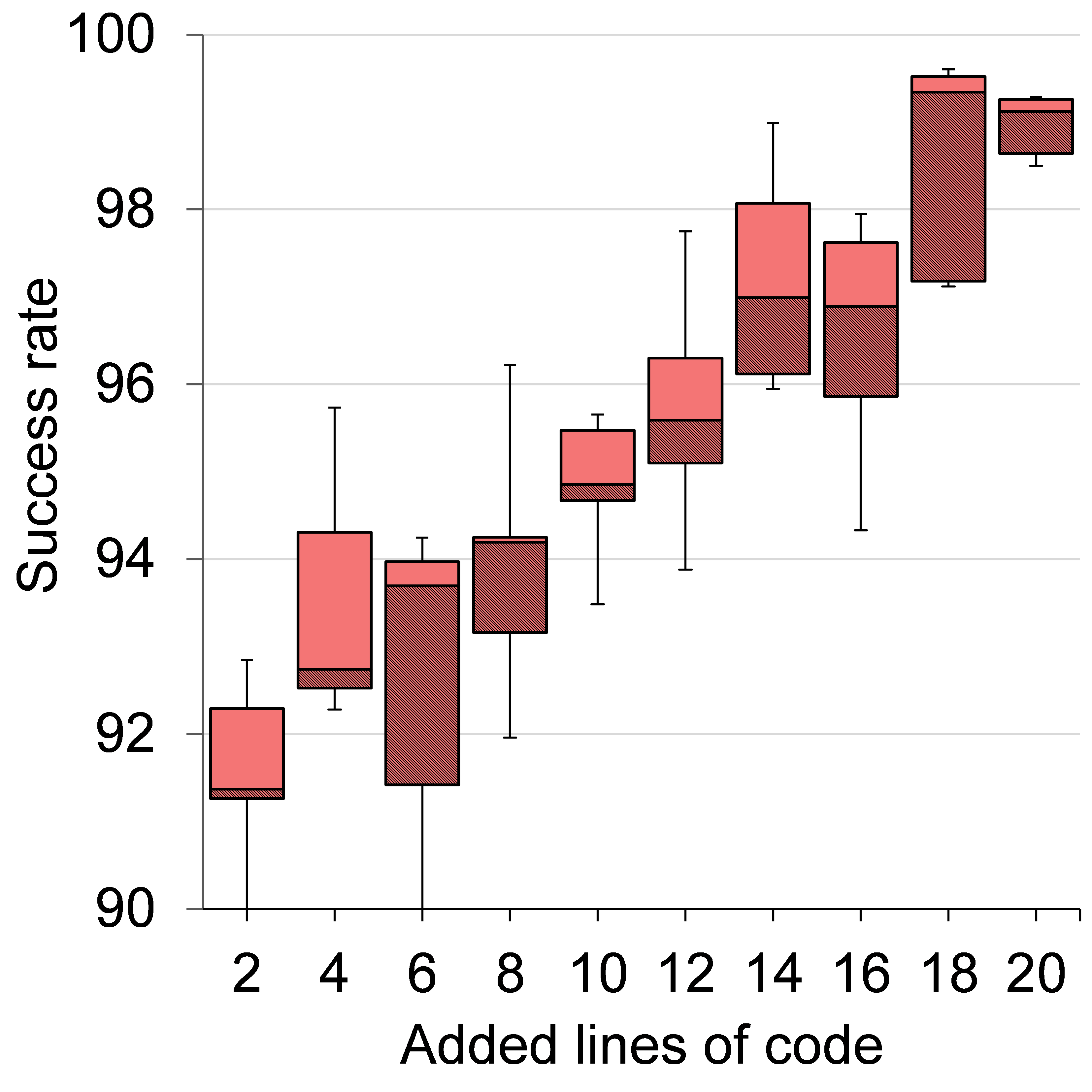}} 
\end{subfigure}
\begin{subfigure}[TFIDF-C-CNN \label{fig:untargeted_TFIDF-C-CNN}]{\includegraphics[width=.155\textwidth]{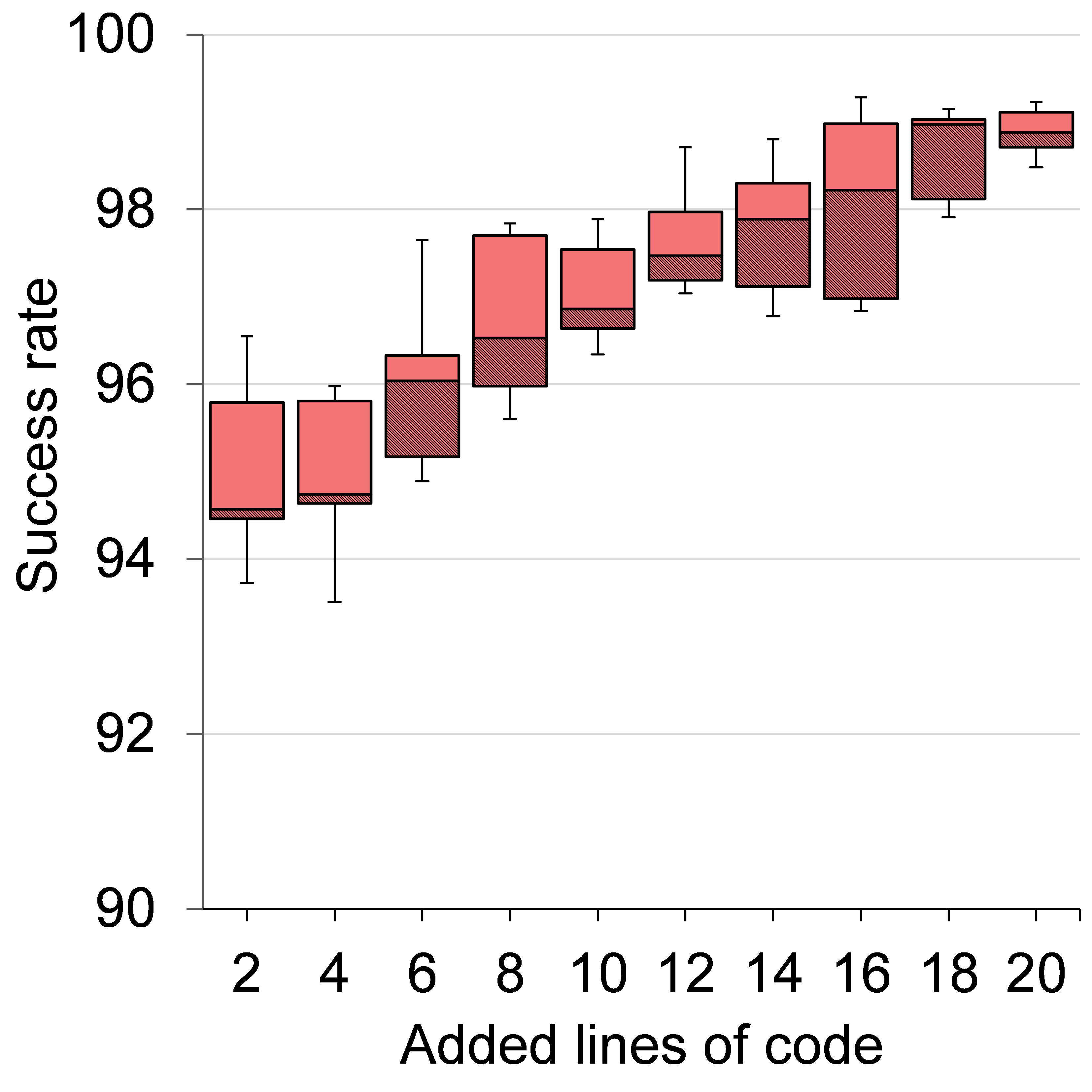}} 
\end{subfigure}
\begin{subfigure}[TFIDF-S-CNN \label{fig:untargeted_TFIDF-S-CNN}]{\includegraphics[width=0.155\textwidth]{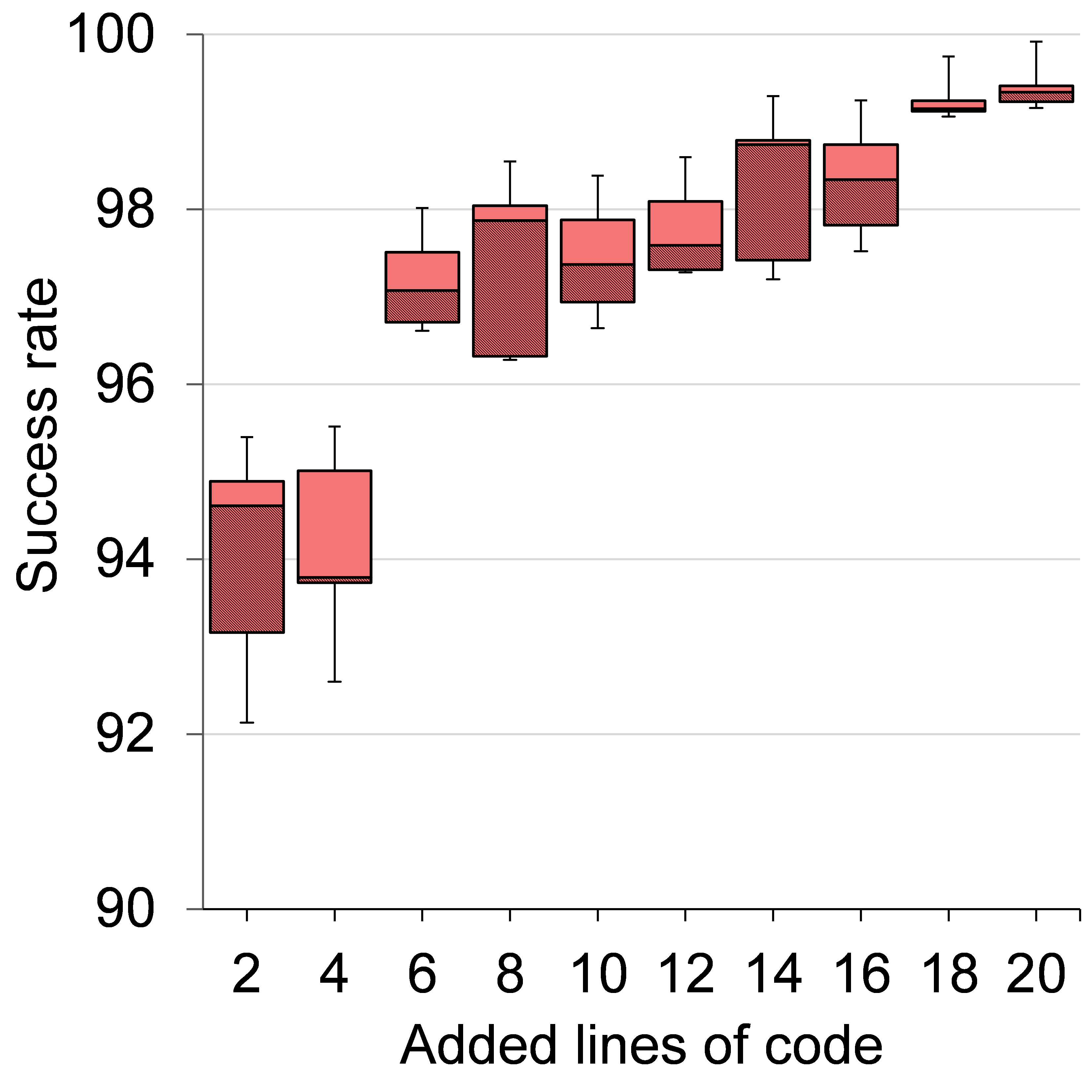}} 
\end{subfigure}
\begin{subfigure}[WE-C-CNN \label{fig:untargeted_WE-C-CNN}]{\includegraphics[width=0.155\textwidth]{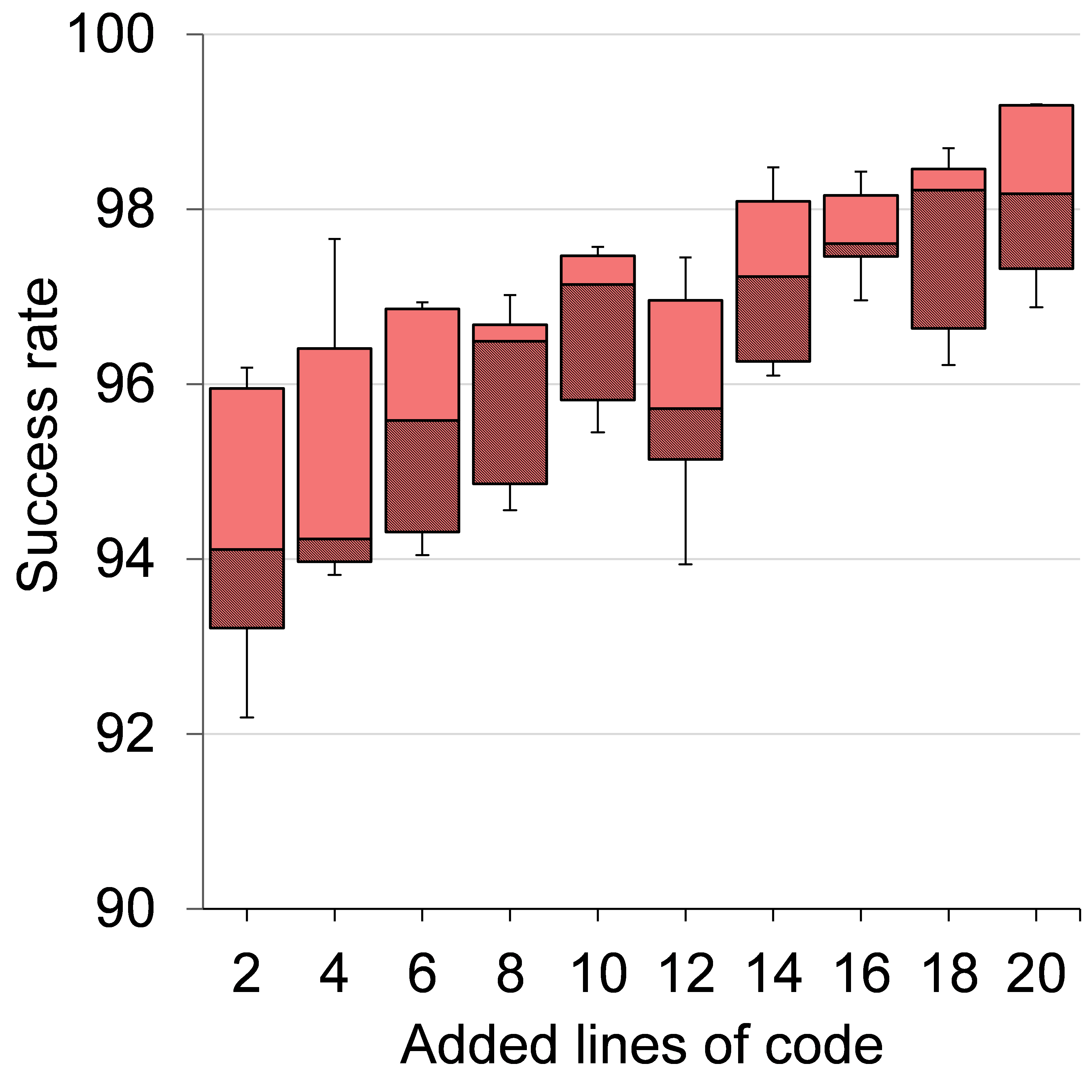}} 
\end{subfigure}
\begin{subfigure}[WE-S-CNN \label{fig:untargeted_WE-S-CNN}]{\includegraphics[width=0.155\textwidth]{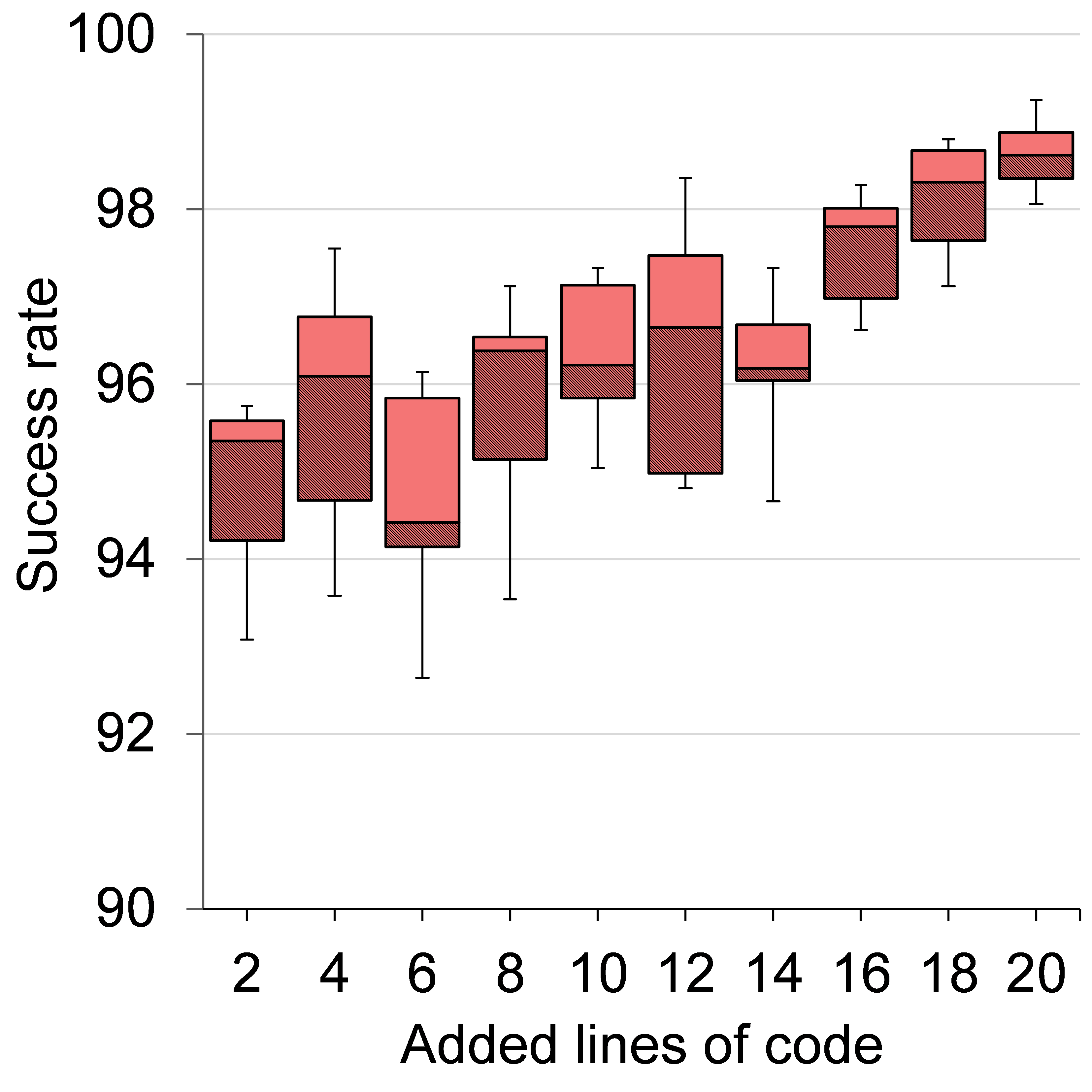}} 
\end{subfigure}\vspace{-3mm}
\caption{{The results of non-targeted attacks using a dataset of 200 programmers. The figure shows the average misidentification rate of different targeted approaches using added perturbations of code lines ranging from 2 to 20.}}
\label{fig:untargeted_attacks_per_line}\vspace{-5mm}
\end{figure*}


\subsection{Experiments: Impact of Attacks} \label{non-targeted_evaluation}

Figure~\ref{fig:untargeted_attack} shows the misidentification rate before and after launching the non-targeted attack. 
The attack produces an average misidentification rate above 98.8\% for all targeted approaches using a dataset of 50 programmers. The rate increases as the dataset size increases, as shown in Figure~\ref{fig:untargeted_attack}.

This misidentification rate is caused by the low confidence levels of identifying programmers when adding code perturbations.
Figure~\ref{fig:ConfidenceReduction} shows the confidence score of the identification models before and after the attack using different datasets.
The figure shows the clear impact of code perturbation on the confidence score of the models: when using a dataset of 200 programmers, the confidence score reaches a high point of 60.69\% and 52.46\% for DL-CAIS and CSFS, respectively, and a low point of 49.27\% for WE-S-CNN.

\BfPara{How the Number of Programmers Impact Misclassification} 
{Figure~\ref{fig:untargeted_attack} shows the average misidentification rate of different methods using different datasets of programmers ranging from 50 to 200. 
Even when the number of programmers is as low as 50, the attribution systems failed to generate the right output since the average achieved misidentification rate was higher than 98.8\%.
The results show that increasing the dataset size increases the misidentification rate. 
A similar trend is observed in reporting the confidence score of models, as the larger the model's output size the more it is affected by the perturbation in terms of output confidence. 
For example, the confidence scores of DL-CAIS and CSFS after the attack are 70.61\% and 64.98\% when using the 50-programmer dataset, and 60.69\% and 52.46\% when using the 200-programmer dataset. 
These scores and the drop in confidence (i.e.,  before and after the attack) are shown in Figure~\ref{fig:ConfidenceReduction}.}

\BfPara{How the Number of Added Code Lines Impact the Success Rate} 
{To explore the impact of perturbation size on the attack success rate, we restricted the implementation of the attack to add a specific number of code lines. The attack starts with adding two lines of code and iteratively probing the target system to a predefined number of access permissions (e.g., in this experiment 20 times). If the attack succeeds, the termination mechanism takes place. Otherwise, the attack generator increases the number of added lines with a step of two. When reaching the predefined limit without succeeding, the attack generator terminates and marks the process as failed.

Figure~\ref{fig:untargeted_attacks_per_line} shows the average misidentification rate achieved when launching the non-targeted attacks with a perturbation of 2 to 20 lines of code using a dataset of 200 programmers. 
Typically, increasing the number of lines affects the overall performance of the attack. For example, the difference in the attack's success rate is ($    99.16 - 94.3 =4.86\%$) when targeting CSFS by adding 2 and 20 lines of code. 
However, this difference becomes less obvious when using larger perturbation than ten lines of code (e.g., $99.41 - 97.62 = 1.79\%$ when targeting TFIDF-S-CNN by adding 10 and 20 lines of code, which are {14.62}\% and {29.23}\% of the average line of codes per file in the dataset, respectively). 
These results confirm that using random code perturbation cripples the identification models' capabilities for identifying programmers. Moreover, even when using the smallest size of perturbation (two lines of code), 
the achieved misidentification rate exceeds 92\% for all approaches when using a dataset of 200 programmers.}

\section{Targeted Attacks}  \label{sec:targeted_attacks}
This section describes the targeted attacks and their impact on the performance of the authorship attribution systems.
 
\subsection{Threat Model 2: Targeted Attack} \label{sec:threatModel2}
{In this setting, the \ours aims to fool the model to predict a specific \textit{target} class (i.e.,  a programmer to be imitated/impersonated).
Targeted attacks can serve multiple adversarial objectives. For example, a common adversarial objective is the \textit{imitation attack}, also referred to as a \textit{impersonation attack}. This kind of attack has serious implications, for example, a malicious programmer can manipulate the code to blame a benign programmer. Another adversarial objective is the \textit{disguise attack} or \textit{evasion attack}, where the \ours impersonates the closest programmer to their style as a disguise. Even though \ours does not target a specific programmer for impersonation, the attack is considered a targeted attack.}

\BfPara{Definition}
The targeted attack aims to maximize the probability of an adversarial code sample to be classified as the targeted class. 
If successful, this attack enables programmers to imitate the style of other programmers. 
Since the identification decision is the $\argmax_k \textit{conf~}(\mathcal{Y}|x)$, the model predictions can be changed based on the distribution of classes.
Using the same definition of $\textit{conf~}(y|x)$, the aim is to minimize the confidence of predicting the right programmer and maximizing the probability of the target.

\BfPara{Adversarial Objective} 
The adversarial goal of targeted attack is to maximize the confidence of the models' predictions towards a target class $\bar{y}$ to lead the model to predict $\bar{y}$.
Similar to the previous threat model, this is conducted by code perturbation $\delta$ on $x$ such as the generated adversarial
code $\bar{x} = x + \delta$ serves the purpose of maximizing the confidence of a target programmer $\bar{y}$. 
Thus, the adversarial objective is formally defined as: 
{\small\[ 
f_{\delta} (x) = \min_{\delta} \{|\delta|~ s.t. ~ \{\textit{conf~}(\bar{y}|\bar{x}) > \{\max_{i}( \textit{conf~}(y_i|\bar{x})) ~\forall y_i \neq \bar{y}\}  \} 
\]}
The attack achieves the objective when the model predicts the targeted programmer to be the author of $\bar{x}$, i.e.,  predicting the $k$-th label as $\bar{y}$ where $\argmax_k \{\textit{conf~}(y_k|\bar{x})\} =  \textit{conf~}(\bar{y}|\bar{x})$.

\BfPara{Adversarial Capabilities} 
{For the targeted attacks, we investigate different adversarial capabilities described as follows.}

\noindent {\bf\em $\bullet$ Targeted Attacks (T1):} 
A \textit{black-box} attack where \ours sends the AE and obtains the model output and scores. 
By iterative probing, \ours optimizes the randomly-generated code perturbations to achieve her goal without any knowledge of the target's coding style.

\noindent \textit{\bf\em $\bullet$ Targeted Attacks (T2):} A \textit{black-box} attack, although \ours has access to two code samples of the target programmer. 
The perturbations are then generated and optimized based on the most representative features of the target programmer.

\noindent {\bf\em $\bullet$ Targeted Attacks (T3):} This attack assumes that \ours has access to code samples of a group of programmers (e.g., ten programmers) and aims to disguise as the closest programmer to her coding style, thus the attack is called \textit{ disguise} or \textit{evasion attack}. Note that the adversarial objective here is not exactly the same as the general objective of the targeted attack, since \ours impersonates the closest programmer to her style to evade identification.
Implementing this attack is easier compared to T1 and T2 due to the flexibility of the imitation options in assigning an adversarial class, especially when the target set is large.

For T3 (disguise/evasion attack),
the attack goal is to maximize the second highest confidence of the model predictions so that $ \textit{conf~}(\bar{y}|\bar{x}) > \textit{conf~}(y|\bar{x})$, where $\bar{y}$ is the closest programmer in the coding style to the adversary.  This does not necessarily mean that the adversarial objective is to predict (specifically) the $\bar{y}$ target since $\argmax_k  \textit{conf~}(y_k|\bar{x})  ~\forall k$ can refer to any class from the target set. 
To select the target, the closest programmer to the adversary in terms of coding style, the adversary needs access to samples from the target set. 

Let the target set of $m$ programmers be $\{y_1, y_2, \dots, y_m\}$, each with $n$ code samples. In our experiments, we assume that the adversary has access to two code samples for each programmer in the target set. For a programmer $y_i$, the closest programmer $\bar{y}$ is the one with the code samples with the shortest Euclidean distance to the samples of $y_i$.  To estimate the Euclidean distance, the adversary represents the accessible samples from the target set using a straightforward $n$-gram model that uses only the occurrences of unigrams.
For each programmer, the adversary obtains the average representation of all samples such that a programmer $y_i$ has the average vector representation
$\text{avg}(x_i) = \frac{1}{n} \text{sum}_{k=1}^{n} (x_i^{(j)} \quad\forall~ j \in \{1, 2 \dots, n\})$, where $n$ is the number of samples for a programmer $y_i$, $x_i^{(j)} \in \mathbb{R}^d$ is the vector representation of the $j$-th sample, $d$ is the dimension of extracted unigrams, and the $\text{sum}$ is the {\em point-wise vector summation operation}.

The closest programmer $\bar{y}$ to a programmer $y$ with samples of $\text{avg}(x)$ is then selected using the shortest distance. Then, $\bar{y} = y_i$, for $y_i$ that satisfies
$\mathsf{min}_{i} \Big(\mathsf{dis}\big(\text{avg}(x) - \text{avg}(x_i)\big)\Big) \forall~ i \in \{1, \dots, m\}$, where $m$ is the number of programmers in the target set, and $\mathsf{dis}$ is the euclidean distance between two average vector representations, defined as:
$\mathsf{dis} (\text{avg}(x), \text{avg}(x_i)) = 
\small{\sqrt{\sum_{j} \Big(\text{avg}(x)^{(j)} - \text{avg}(x_i)^{(j)}\Big)^2 }} \quad \forall~ j \in \{1, \dots, d\}$.

After defining the nearest programmer $\bar{y}$, the adversary retrieves the most influential features using the order of the features to be the source for the perturbation. 
Similar to the imitation attack, this attack is conducted by adding code perturbation $\delta$ to the input code $x$ so that the generated adversarial  
code $\bar{x} = x + \delta$ maximizes the confidence of predicting the programmer $\bar{y}$. 


\begin{figure}[t]
\centering
\includegraphics[width=0.3\textwidth]{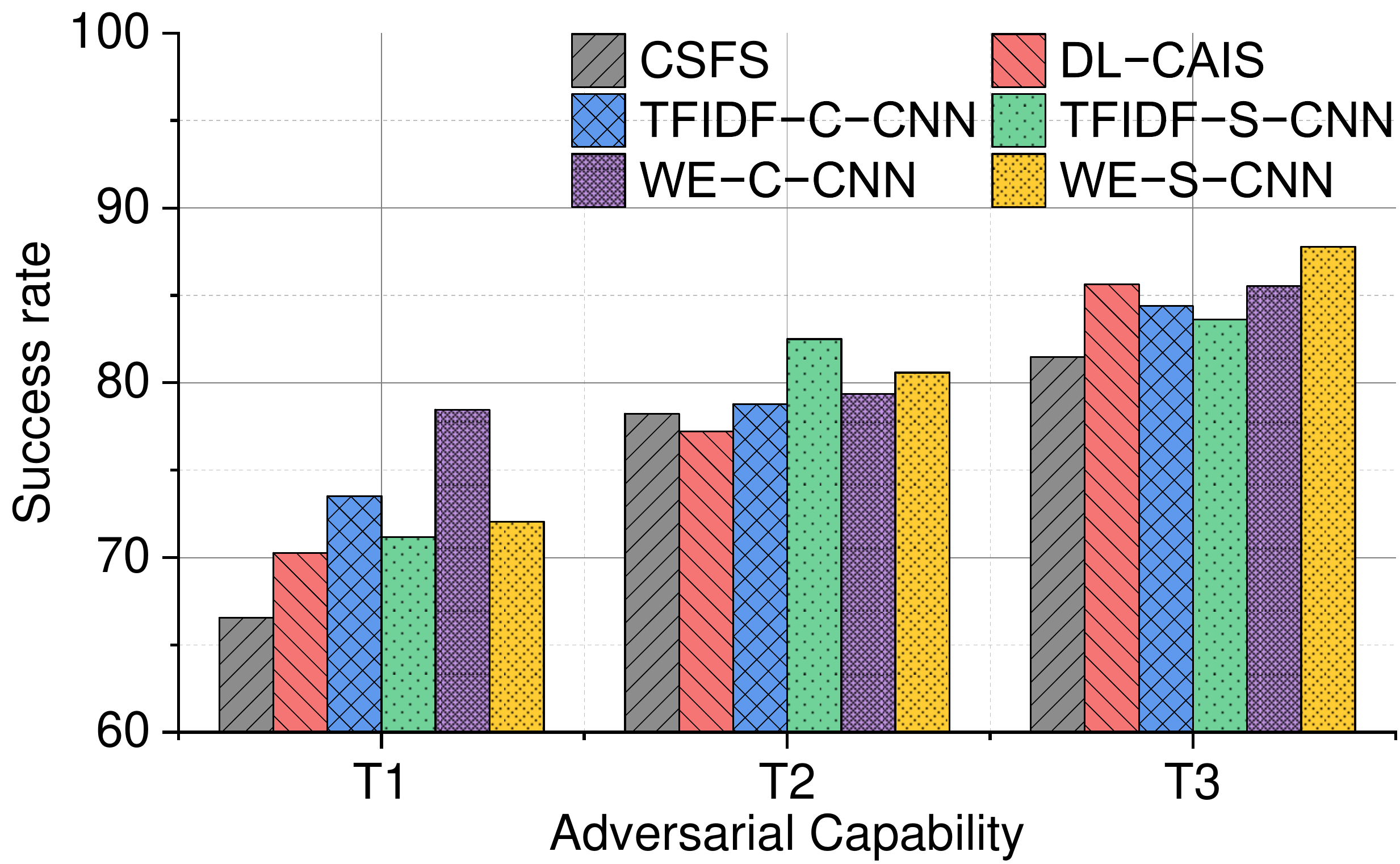}\vspace{-3mm} 
\caption{{The results of the targeted attack: the success rate  of attacks on the targeted approaches categorized by different threat models. The results show the increase in the success rate is associated with adversarial knowledge.}}
\label{fig:average_targeted_attacks}\vspace{-4mm}
\end{figure}

\subsection{Dataset and Evaluation Metrics} \label{sec:dataset}

\BfPara{Our Dataset}
For this threat model, we use a subset of the same dataset from the previous experiment. 
To evaluate this attack, we randomly selected a subset of 100 programmers to simulate different experimental settings.
In this experiment, we marked each programmer in our subset dataset of 100 programmers to be subject to the imitation attack while using all other programmers to imitate the coding style of that target programmer. 
Since our dataset includes nine code samples per programmer, we generated ($9 \times 99 = 891$) adversarial code samples for each subject programmer, i.e.,  we generating 89,100 adversarial code samples for each experimental setting considering the 100 programs we included. 

One consideration we took into account when choosing the random 100 programmers is that each selected programmer should have more than 11 samples in the GCJ dataset. This constraint is to allow the implementation of the targeted attacks T2 using the same dataset for consistency. We emphasize that T2 assumes knowledge of two samples.

\BfPara{Attack Evaluation Metrics}
We evaluate the targeted attack by its success rate as the proportion of the correctly classified adversarial samples to the targeted programmer with respect to the overall attempts. 
For a targeted class $\bar{y}$, the attack succeeds when the model predicts $\bar{y}$, i.e.,  the model outputs $y_k$ such that $\argmax_k P(\mathcal{Y}|\bar{x}_i)$ refers to $\bar{y}$. 
The success rate is calculated as:
$
\frac{1}{n \times m} \sum_{j}^{m} \sum_{i}^{n}  I(y_k = \bar{y})
$, where $y_k$ is the predicted class at the $k$-th position of the target set (i.e.,  $\argmax_k P(\mathcal{Y}|\bar{x}_i)$).

For T3, i.e.,  evasion attack,
the attack success rate is defined by the rate of correctly misidentifying a specific programmer to all presented code samples by that programmer, i.e.,  similar to the misidentification rate. 
However, we adopted a targeted attack where an adversarial code sample $\bar{x}_i$ is attributed to the closest programmer $\bar{y}$ with respect to the original programmer $y_i$. Imitating the closest programmer is the reason for assuming this scenario as a targeted attack, and it is evaluated as such.

\begin{figure*}[t]
\centering
\begin{minipage}{.96\textwidth}%
	\centering
	\subfigure[CSFS \cite{Caliskan-Islam:2015}]{%
		\includegraphics[width=0.33\textwidth]{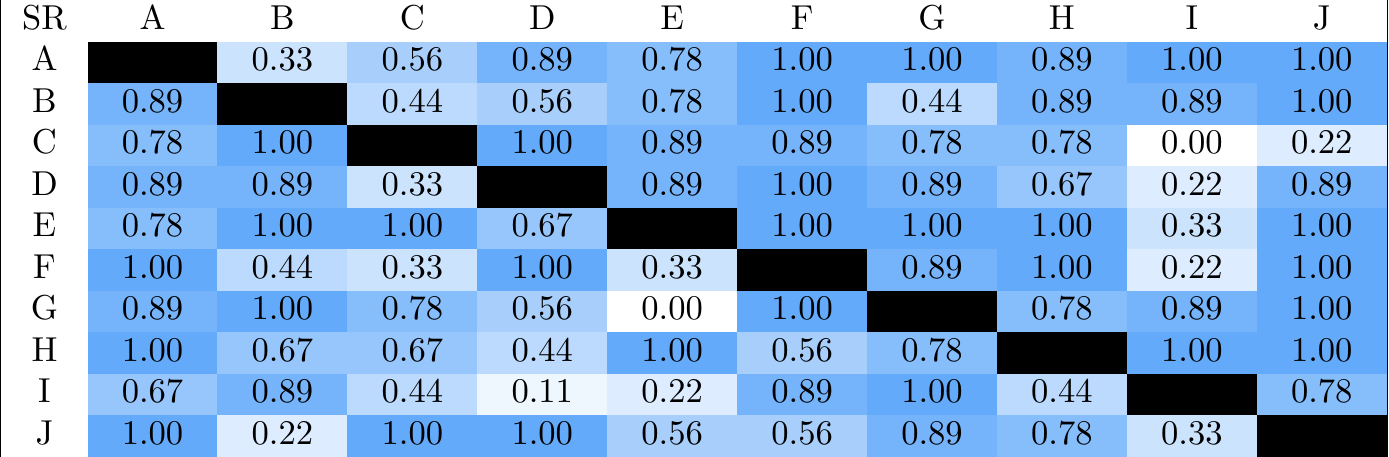}%
		\label{fig:targeted_CSFS}%
	}%
	\subfigure[DL-CAIS \cite{abuhamad2018large}]{%
		\includegraphics[width=0.33\textwidth]{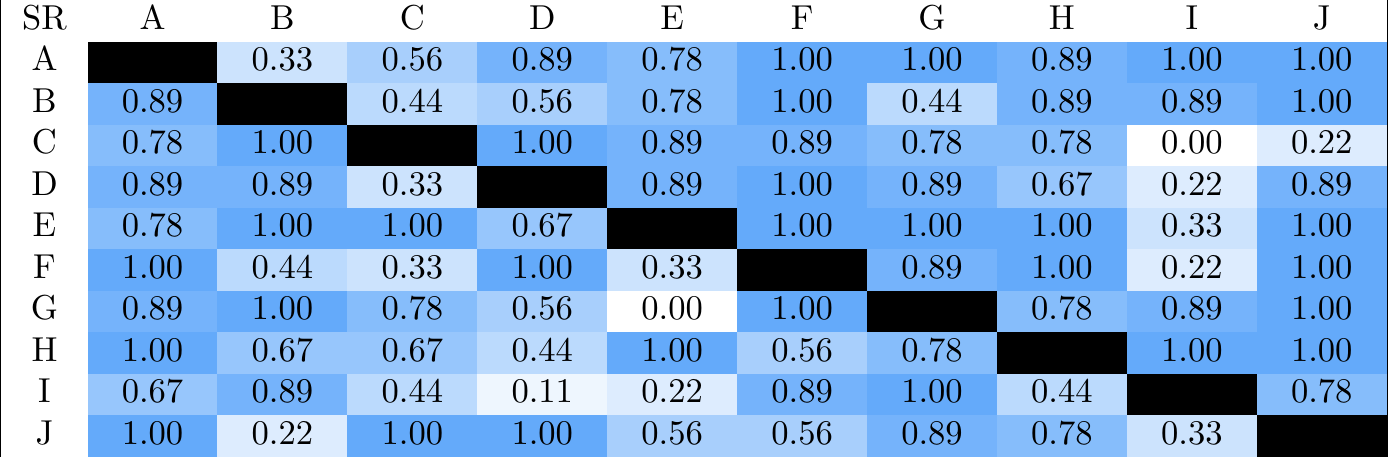}%
		\label{fig:targeted_DL-CAIS}%
	}%
	\subfigure[TFIDF-C-CNN \cite{abuhamad2019code}]{%
		\includegraphics[width=0.33\textwidth]{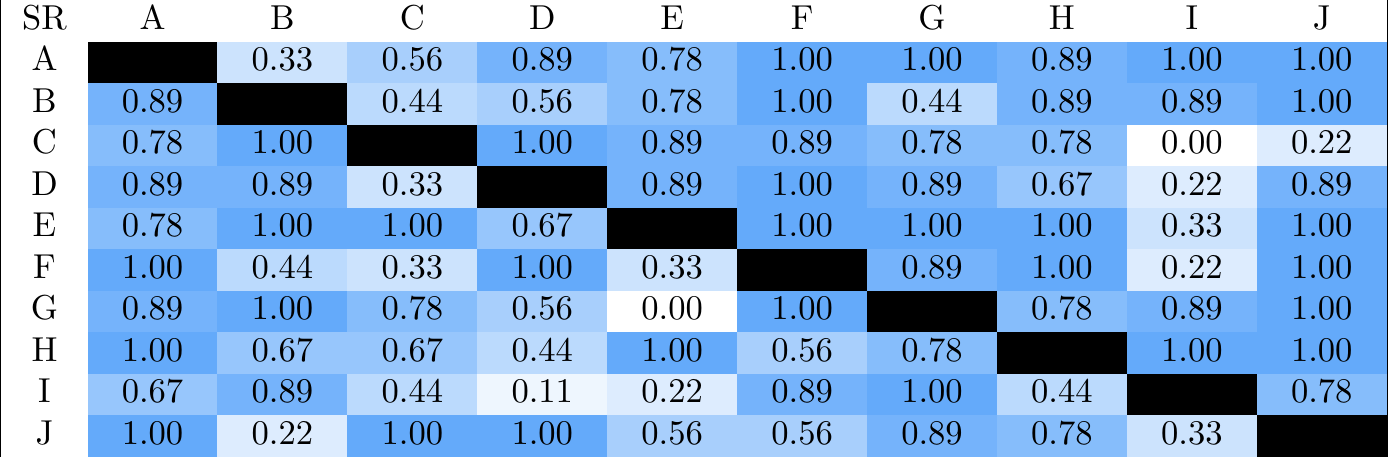}%
		\label{fig:targeted_TFIDF-C-CNN}%
	}%
	\\
	\subfigure[TFIDF-S-CNN \cite{abuhamad2019code}]{%
		\includegraphics[width=0.33\textwidth]{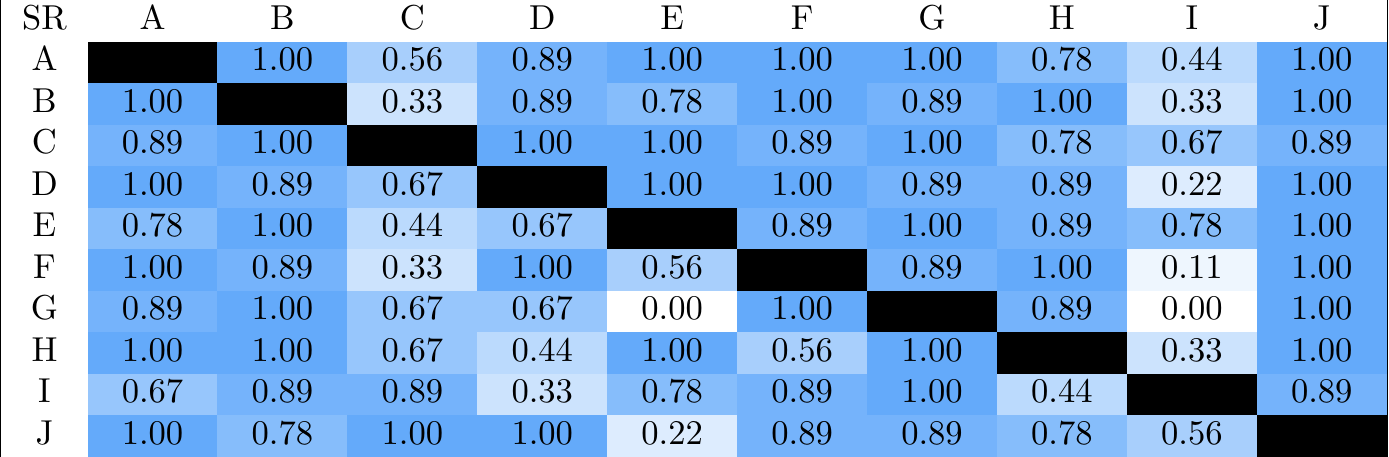}%
		\label{fig:targeted_TFIDF-S-CNN}%
	}%
	\subfigure[WE-C-CNN \cite{abuhamad2019code}]{%
		\includegraphics[width=0.33\textwidth]{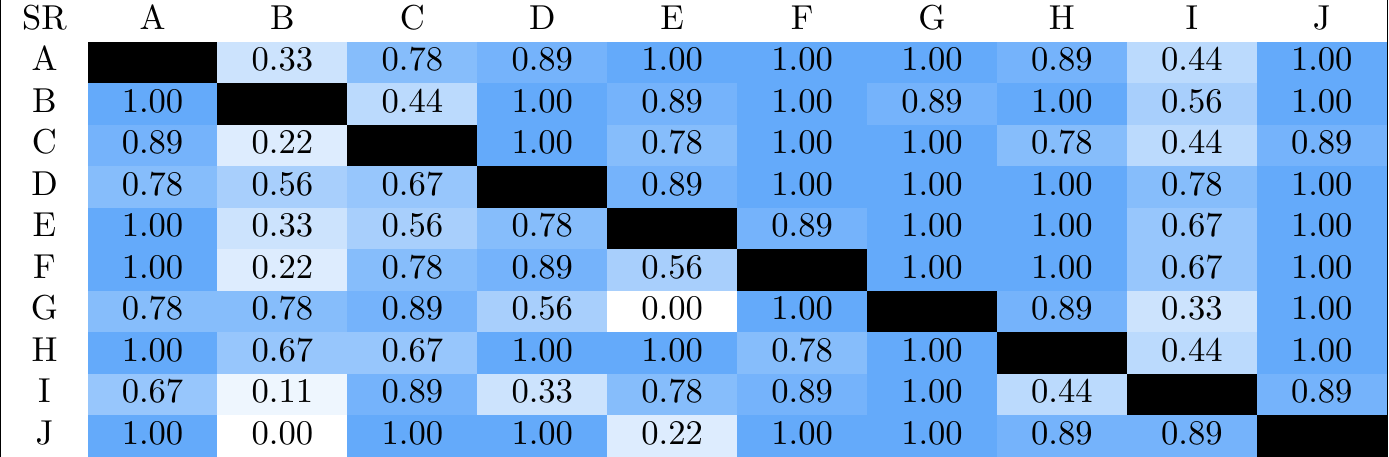}%
		\label{fig:targeted_WE-C-CNN}%
	}%
	\subfigure[WE-S-CNN \cite{abuhamad2019code}]{%
		\includegraphics[width=0.33\textwidth]{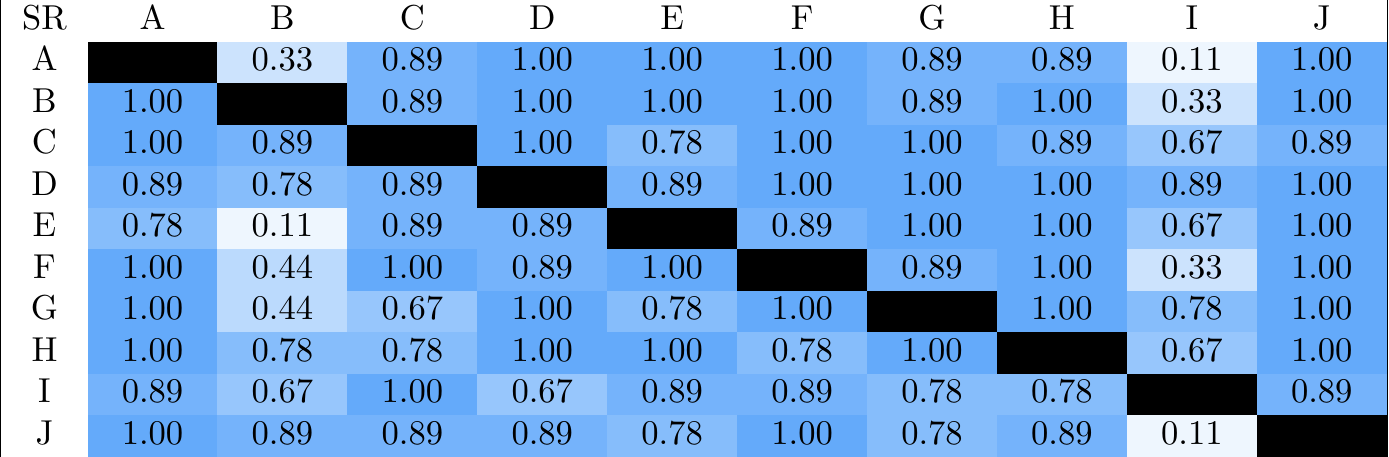}%
		\label{fig:targeted_WE-S-CNN}%
	}%

	\end{minipage}\vspace{-3mm}
	\caption{{Matrix representation of the targeted attack T1 for six authorship attribution systems. Each cell indicates the success rate of the attack for ten programmer pairs using nine code files per programmer.}}
	\label{fig:targeted_maps}\vspace{-5mm}
\end{figure*}

\subsection{Experimental Results: The Impact of the Attacks} \label{targeted_evaluation}

The impact of the different targeted attacks on the performance of the authorship attribution systems is shown in Figure~\ref{fig:average_targeted_attacks}, and in the following we elaborate on those results.

\BfPara{Targeted Attack (T1)}
{Figure~\ref{fig:average_targeted_attacks} shows that
the success rate is considerably high with more than 66.55\% success rate for all systems under the T1 attack. The average success rates are 66.55\%, 70.26\%, 73.51\%, 71.16\%, 78.45\%, and 72.05\% for CSFS, DL-CAIS, TFIDF-C-CNN, TFIDF-S-CNN, WE-C-CNN, and WE-S-CNN, respectively. 

Figure~\ref{fig:targeted_maps} shows the results of ten random programmers as a matrix. Each cell represents the success rate of targeted attacks where each programmer attempts to imitate another programmer in implementing nine programming challenges. The figure shows that most programmers can be imitated and their coding style can be manipulated to match other programmers'. This is true for all targeted systems, with more success rate on the CNN-based systems. The figure also shows that some programmers are more difficult to imitate than others, e.g., programmer I in Figure~\ref{fig:targeted_TFIDF-S-CNN} using TFIDF-S-CNN and programmer B in Figure~\ref{fig:targeted_WE-C-CNN} using WE-C-CNN.}

\BfPara{Targeted Attack (T2)}
{Given more adversarial knowledge, i.e.,  the adversary has access to two samples of the target programmer, the results show an increased success rate for the attacks.
This scenario still assumes a \textit{black-box} attack since the adversary has no knowledge or access to the system's internal components. The success rate of this attack exceeds 77\% for all targeted systems. 
Figure~\ref{fig:average_targeted_attacks} shows that the average success rates are 78.23\%, 77.22\%, 78.79\%, 82.49, 79.35\%, and 80.58\% for CSFS, DL-CAIS, TFIDF-C-CNN, TFIDF-S-CNN, WE-C-CNN, and WE-S-CNN, respectively.} 

These results show the adversary's knowledge contributed significantly to the success rate of the targeted attacks. Having access to code samples from the target has enabled the optimization of adversarial code examples by considering the perturbations that benefited from the actual stylometry features of the target.  This level of knowledge contributed to an increase of 11.68\%, 6.96\%, 7.63\%, 11.33\%, 0.9\%, and 8.53\% in the attack success rate targeting CSFS, DL-CAIS, TFIDF-C-CNN, TFIDF-S-CNN, WE-C-CNN, and WE-S-CNN, respectively.

\BfPara{Targeted Attack (T3)} This model extends the knowledge of the adversary to include access to samples of a subset of programmers to imitate the closest one among them.  This capability allows for a higher success rate in comparison to T1 and T2. We note that the goal of the adversary in T3 is different from that of T1 or T2 (see Section \ref{sec:threatModel2}).
Figure~\ref{fig:average_targeted_attacks} shows that the average success rates are 81.48\%, 85.63\%, 84.40\%, 83.61, 85.52\%, and 87.77\% for CSFS, DL-CAIS, TFIDF-C-CNN, TFIDF-S-CNN, WE-C-CNN, and WE-S-CNN, respectively.  The results show  clearly that the adversary has a higher success chance in imitating one programmer in a group of programmers with access to code samples from those programmers.

\begin{figure}[t]
\centering
\begin{subfigure}[T1 \label{fig:targeted_attacks_per_line-T1}]{\includegraphics[width=0.145\textwidth]{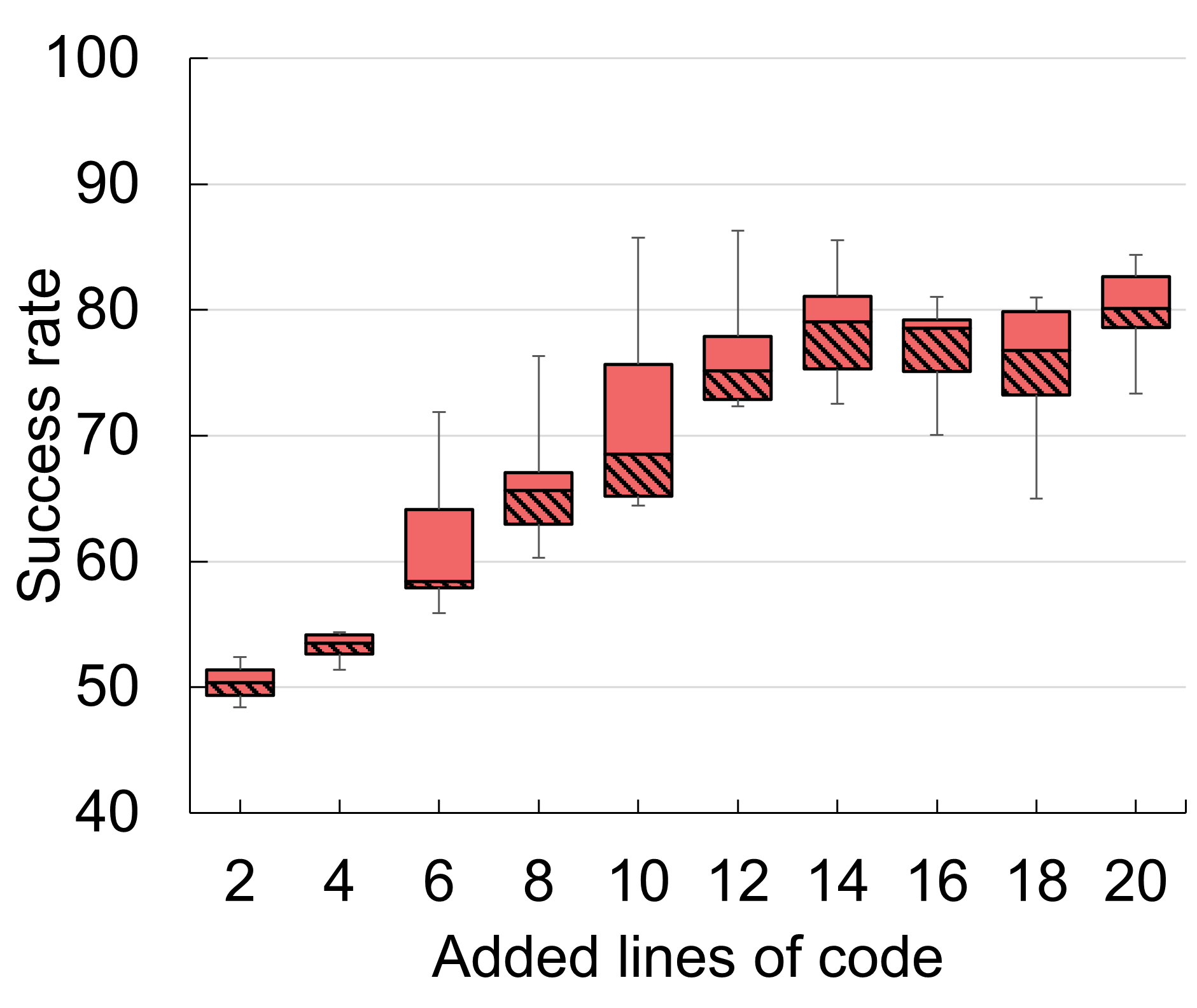}} 
\end{subfigure}~
\begin{subfigure}[T2 \label{fig:targeted_attacks_per_line-T2}]{\includegraphics[width=0.145\textwidth]{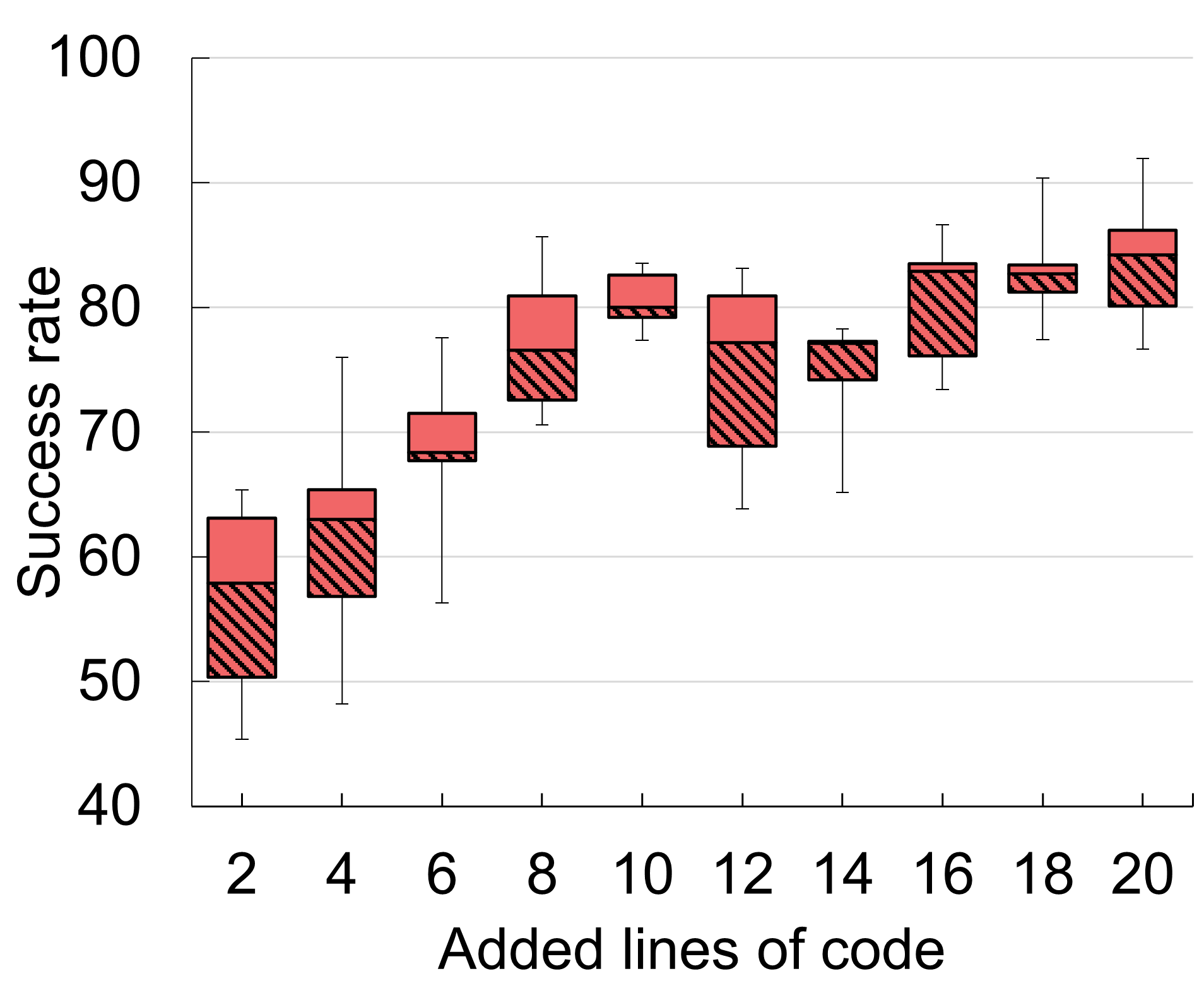}}
\end{subfigure}~
\begin{subfigure}[T3\label{fig:targeted_attacks_per_line-T3}]{\includegraphics[width=0.145\textwidth]{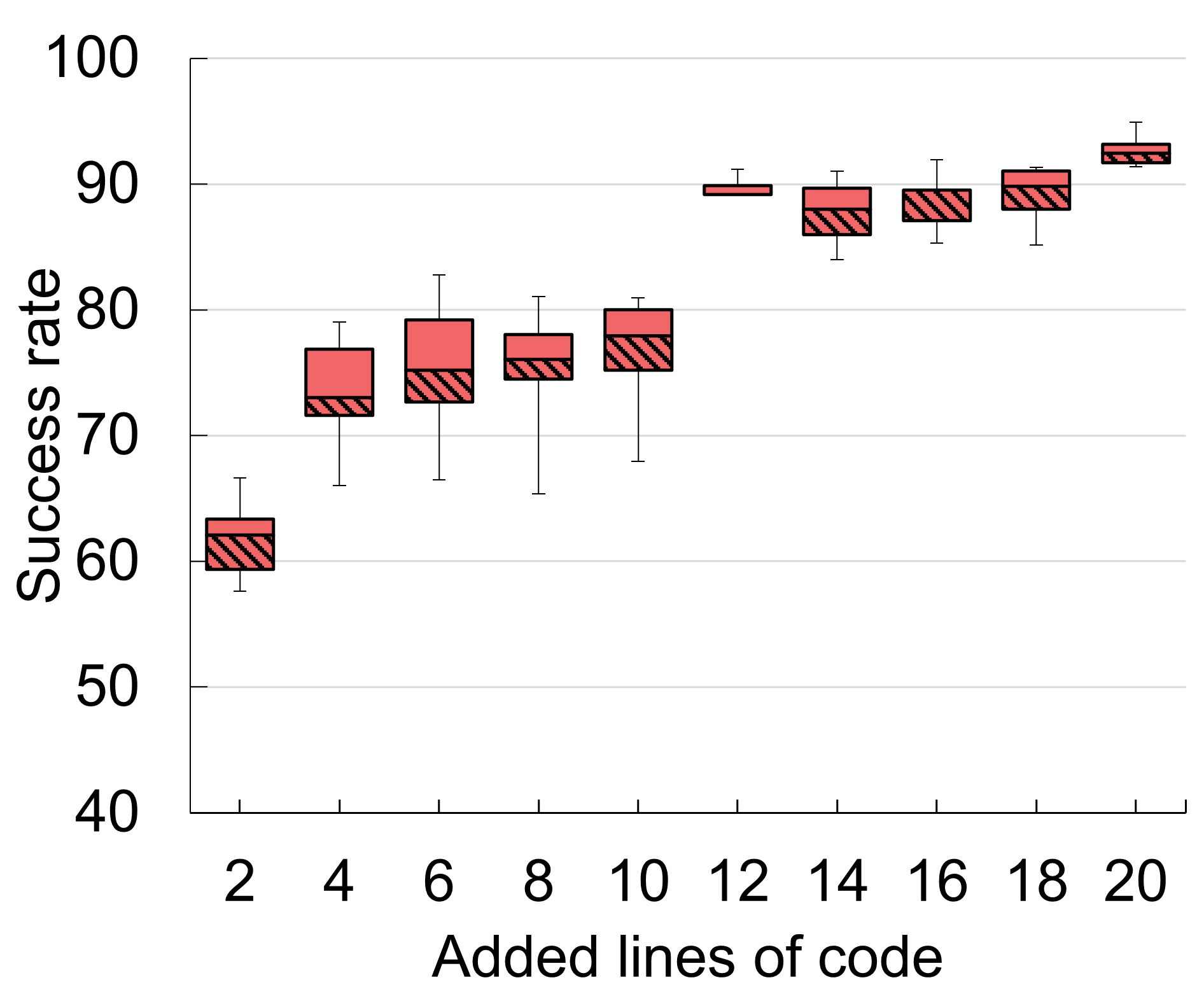}}
\end{subfigure}

\caption{{The results of different targeted attacks using a dataset of 100 programmers. The figure shows the average success rate of the targeted attacks using added perturbations of code lines ranging from 2 to 20.}} 
\label{fig:targeted_attacks_per_line}\vspace{-3mm}
\end{figure}

\BfPara{Impact of Added Code Lines} We implemented the attack with a specific number of code lines. 
Similar to the settings adopted for non-targeted attacks, the attack generator increases the number of added lines with a step of two and iteratively accessing the target system to a predefined number of access permissions each step (e.g., in this experiment 50 times). 
Figure~\ref{fig:targeted_attacks_per_line} shows the average success rate of different attack scenarios using different lines of code as perturbations. 

The results show that code perturbation can lead to a high success rate of targeted attacks when adding as few as two lines of code. For example, the average success rates achieved under T1, T2, and T3 are 50.37\%, 55.37\%, and 61.12\%, respectively, for all targeted systems when using only two lines of code as a perturbation. When increasing the perturbation size to 20 lines of code, the success rate reaches 83.86\%, 83.58\%, and  92.03\% for the threat models T1, T2, and T3, respectively.
These results demonstrate that most authorship attribution systems are susceptible to adversarial perturbations and can be highly affected even with small changes in the input file.

\begin{figure*}[t]
\centering
\begin{subfigure}[Original \label{fig:OriginalPCA}]{\includegraphics[width=0.24\textwidth]{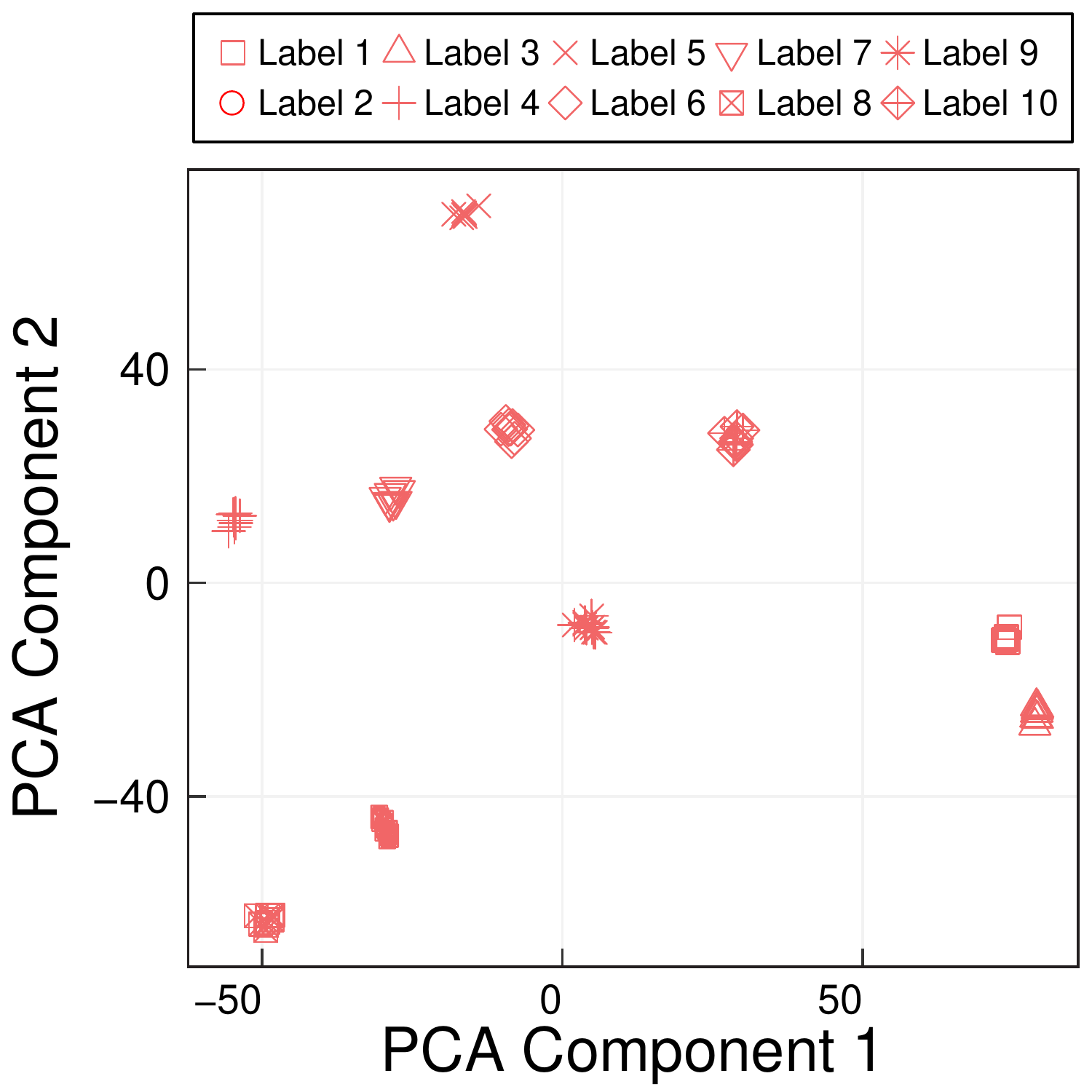}}
\end{subfigure}
\begin{subfigure}[Non-targeted Attack \label{fig:ConfidencePCA}]{\includegraphics[width=0.24\textwidth]{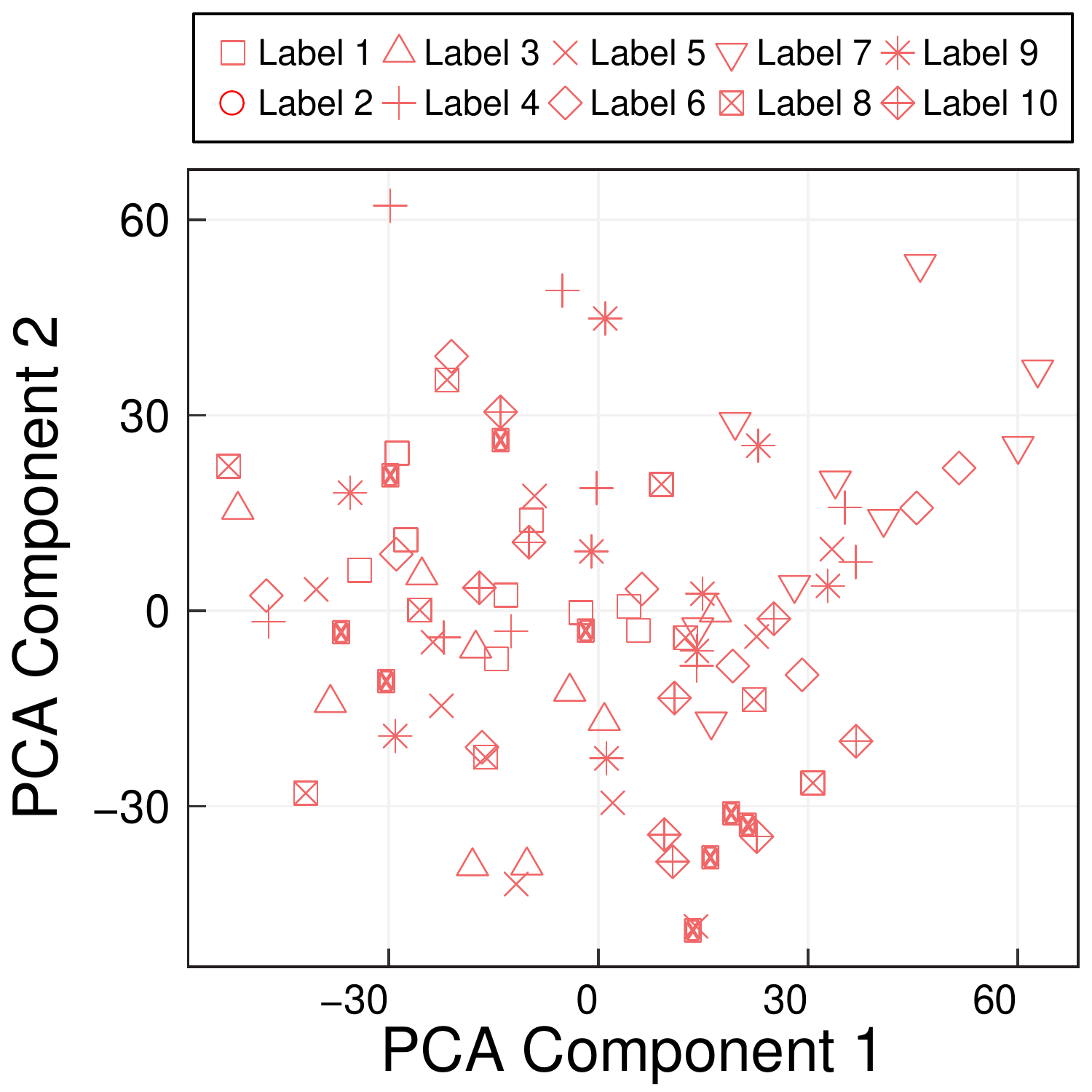}}
\end{subfigure}
\begin{subfigure}[Targeted Attack (T1) \label{fig:imitationPCA}]{\includegraphics[width=0.24\textwidth]{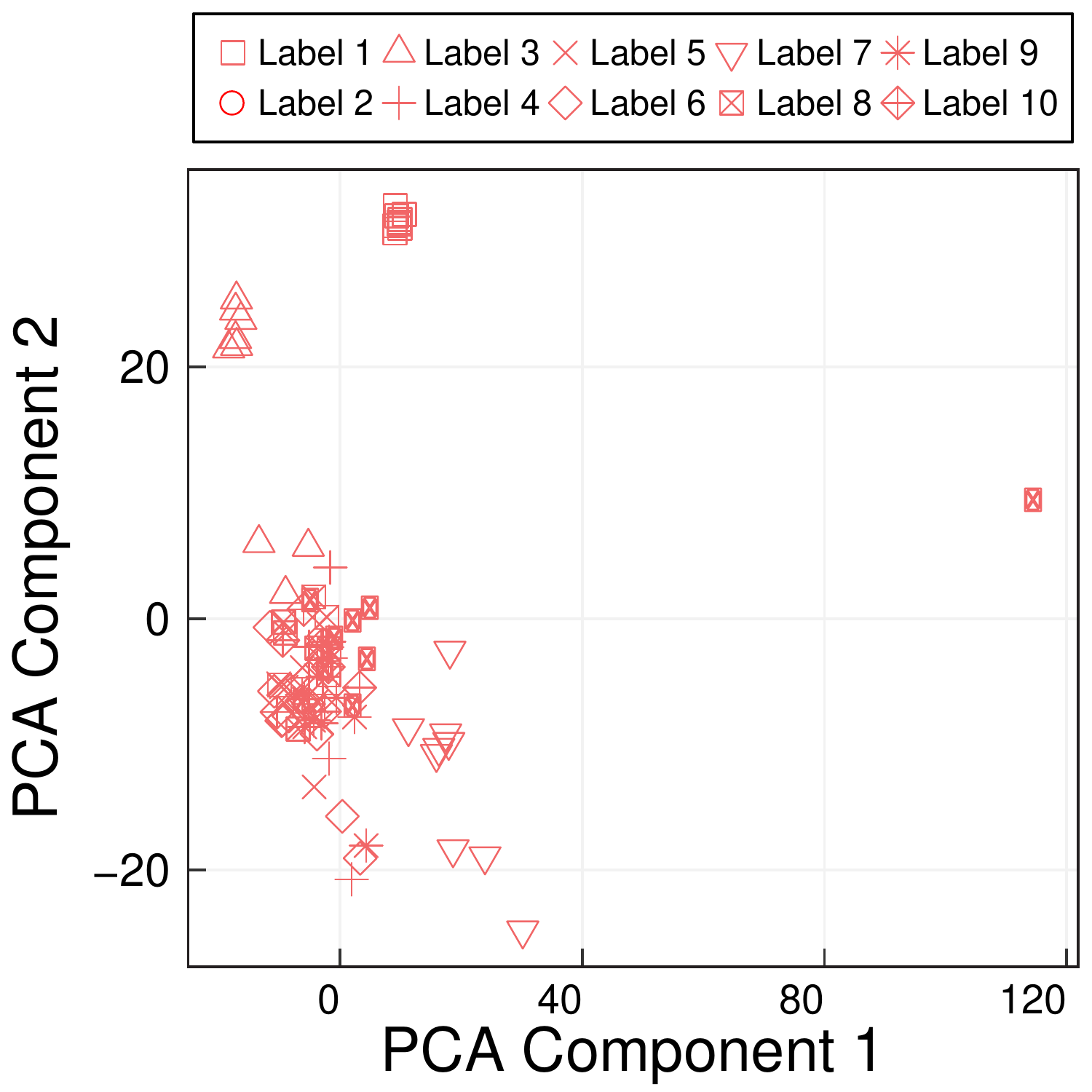}}
\end{subfigure}
\begin{subfigure}[Targeted Attack (T3) \label{fig:evasionPCA}]{\includegraphics[width=0.24\textwidth]{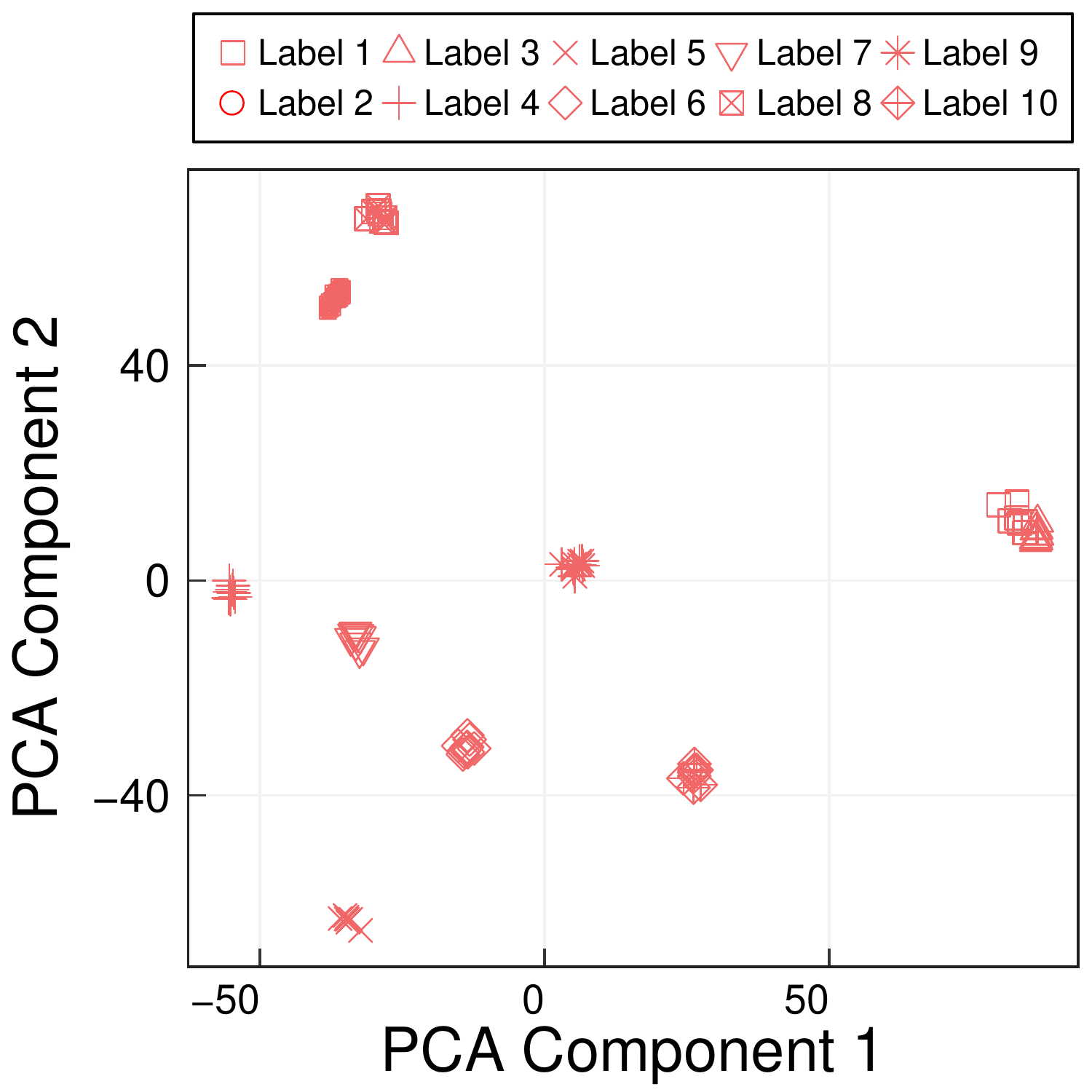}}
\end{subfigure}\vspace{-5mm}
\caption{The PCA visualization of authorship attributions of 10 programmers with nine code samples. The figure shows the original attributions along with the effects of different adversarial attacks on the generated attributions using DL-CAIS.} 
\label{fig:PCAVisualization}\vspace{-3mm}
\end{figure*}

\section{Discussion} \label{sec:discussion}
We now explore the effects of adversarial perturbations on the authorship attributes against different attacks. Second, we show the magnitude of perturbations when adding code blocks to the original code. We compare the performance of \ours{} to related work, and finally, we list the limitations.

\begin{figure}[t]
\begin{minipage}{.24\textwidth}
\centering
\includegraphics[width=0.99\textwidth]{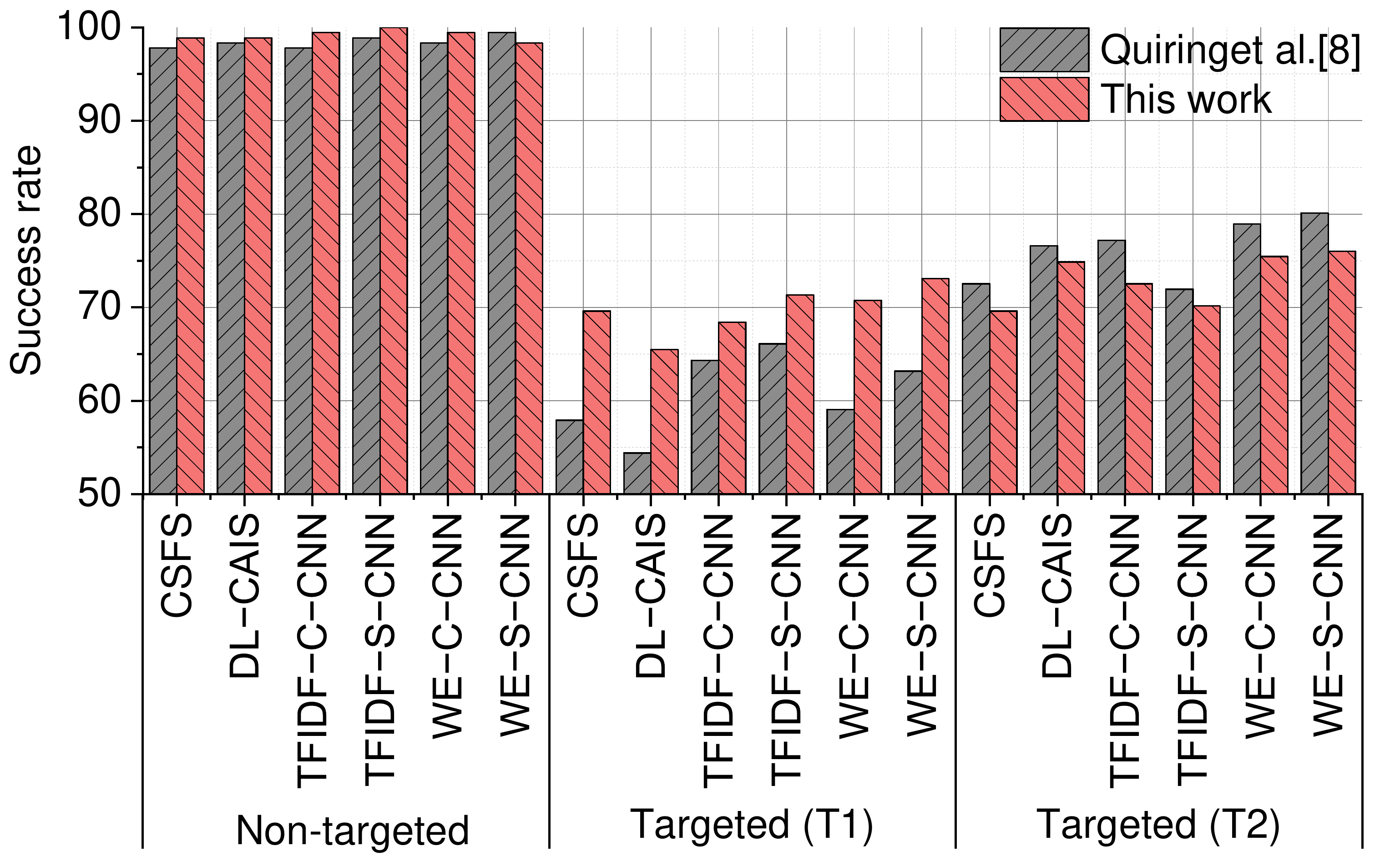}\vspace{-3mm} 
\caption{{Comparison with the work of Quiring \etal~\cite{quiring2019misleading} using three adversarial settings. The results are obtained using a dataset of 20 programmers and access to the model 20 times.}}
\label{figure:Comparison}\vspace{-3mm}
\end{minipage}~
\begin{minipage}{.24\textwidth}
\centering 
\includegraphics[width=0.99\textwidth]{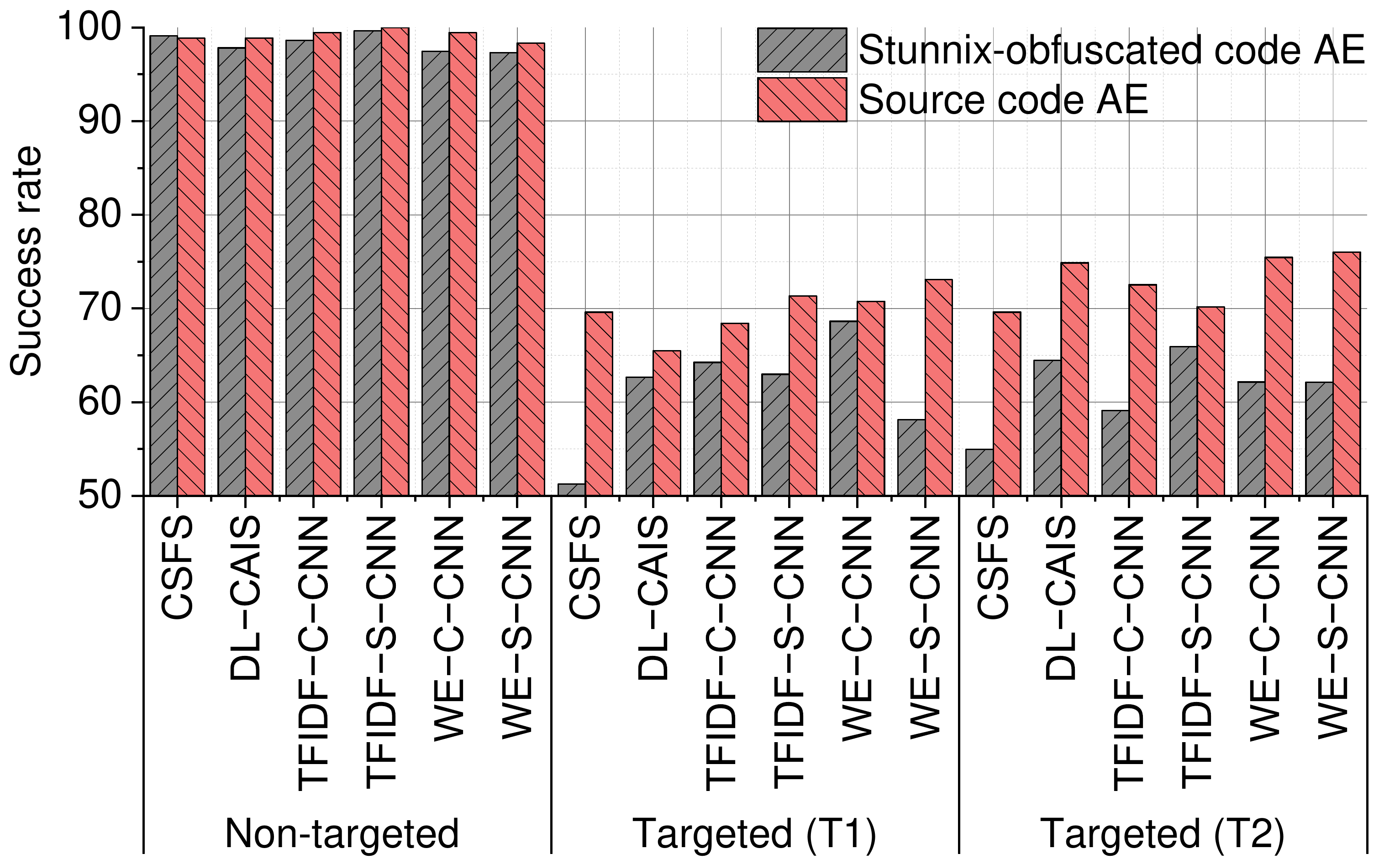}
\caption{The results of different attacks using the original and the obfuscated adversarial code examples. The results obtained using a dataset of 20 programmers.}
\label{figure:obfuscation_stunnix}\vspace{-3mm}
\end{minipage}
\end{figure}

\subsection{Adversarial Authorship Attributions}
We show that code authorship attribution systems can be vulnerable to adversarial attacks. 
For all attacks, the authorship attributes are greatly affected by the smallest size of perturbations. 
This effect can be shown in the PCA visualization of the authorship attributes of code samples generated by different adversarial settings, as shown visually in
Figure~\ref{fig:PCAVisualization} for code samples of ten programmers with nine code samples each in various adversarial settings. For this visualization, we used the DL-CAIS system to demonstrate the effect of different attacks.
Figure~\ref{fig:OriginalPCA} shows the original code representations of the ten programmers with nine code samples. It is clear that such authorship attributes are the reason for the accurate identification process. 
However, when introducing adversarial code examples as in Figure~\ref{fig:ConfidencePCA} during the non-targeted attack, the authorship features become extremely scattered in the feature space so it is difficult to establish decision boundaries, and hence the high misidentification rate. 
Figure~\ref{fig:imitationPCA} shows a visualization of targeted attack T1, where nine programmers attempt to imitate one label (i.e.,  {\em Label 10} in this figure), which leads to partial success as some programmers are still resilient to imitate other programmers.  The figure shows also that most code samples are within the same proximity as the code samples of the targeted programmer. However, some programmers have very distinct coding,for example, {\em Label 1} has code samples that are not affected by the targeted perturbations. Other programmers are partially affected, such as ({\em Label 3} and {\em Label 7}). 
In Figure~\ref{fig:evasionPCA}, we show the effect of the targeted attack T3 (i.e., evasion attack), in which the adversary aims to imitate the closest programming style to the adversary's. In this figure, {\em Label 1} imitates the coding style of {\em Label 3} as their coding style is the closest among others as also can be seen in Figure~\ref{fig:OriginalPCA}. We explained this attack as if {\em Label 1} is using {\em Label 3} as a disguise to evade identification.

\subsection{Comparison with Other Methods} \label{sec:Comparison}

{We conducted a comparison with the work of Quiring \etal~\cite{quiring2019misleading} using three different adversarial settings. The experiments included the six targeted systems and a dataset of 20 programmers. We followed the implementation details of the original work of \cite{quiring2019misleading} and limited access to the identification model to 20 times. If the attack generator fails to achieve the adversarial goal by the 20-th iteration, it is considered a failed attempt. Figure \ref{figure:Comparison} shows that the adversarial perturbation is as effective as code transformation methods, however considerably simpler. For the non-targeted attacks, both this work and the work of \cite{quiring2019misleading} achieved more than a 97\% success rate, indicating the vulnerability of current code authorship attribution methods to adversarial attacks. 

Under T1, \ours{} has a higher attack success rate due to the flexibility in the added perturbation that shifts attributes to the target output. However, the method by \cite{quiring2019misleading} performs better under T2, where the adversary has access to two samples of the target. This is due to the supported template transformations that help in achieving the adversarial objective. }

\subsection{Obfuscated Adversarial Code Examples}
Code obfuscation is one way to thwart static analysis and elimination of the perturbation in the adversarial code examples by figuring out unreachable code blocks.  
Assuming a stronger authorship attribution pipeline that operates on an obfuscated input domain, we experimented with obfuscated  examples. For this experiment, we used a dataset of 20 C++ programmers with nine files each (the same dataset in Section \ref{sec:Comparison}).
We obfuscated the samples using Stunnix~\cite{stunnix}, a popular code-to-code C/C++ obfuscator that gives the code a cryptic appearance while preserving the functionality. Previous code authorship attribution studies have used Stunnix obfuscator to demonstrate the validity of working in an obfuscated domain and produced high identification accuracy~\cite{abuhamad2018large,Caliskan-Islam:2015}.

\BfPara{Assumptions} In the following experiment, we assume that the attribution system can operate in the obfuscated domain, and some models are trained and tested using the obfuscated samples. The adversary aims to hide the code perturbations using the obfuscation tool, knowing that the authorship attribution system can recognize the obfuscation tool and assign the sample to the models trained on the obfuscated domain.  

\BfPara{Identification of Obfuscated Samples} Establishing baseline models that operate in an obfuscated domain, we evaluated the six targeted systems on a dataset of Stunnix-obfuscated code of 20 programmers. The 9-fold cross-validation accuracy was 97.64\%, 95.43\%, 94.79\%, 98.19, 98.82\%, and 96.18\% for CSFS, DL-CAIS, TFIDF-C-CNN, TFIDF-S-CNN, WE-C-CNN, and WE-S-CNN, respectively.

\BfPara{Obfuscated Adversarial Code Examples}
\ours{} follows the same method of generating adversarial code examples, as described in the different threat models (T1-T3), and obfuscates the adversarial code to produced the obfuscated adversarial example. The obfuscated code samples are then submitted to the authorship attribution system to identify authors. We limit the adversary's access to the identification model to 20 times to achieve the adversarial goal. 

Figure \ref{figure:obfuscation_stunnix} shows \ours{}'s success rate of different attacks in the obfuscated domain. Compared to the attacks on the source code domain, the obfuscated code samples achieved a similar success rate in the non-targeted attacks against all targeted systems. For targeted attacks using T1 and T2 threat models, the success rate degrades slightly due to the difficulty of optimizing the adversarial code example at the source code level using the feedback of the model on the obfuscated sample. Moreover, the perturbations on the source level (before the obfuscation) do not guarantee a shift in the feature space of the obfuscated sample (after obfuscation) toward a certain target. However, the attacks exceed the success rate of 51.24\% and 54.96\% for targeted attacks T1 and T2, respectively.

\subsection{The Impact of the Perturbation Size}

We use $\ell_p$ to measure the magnitude of perturbation by $p\mbox{--}norm$
distance as:
$\left \| \delta \right \|_p = \left ( \sum_{i=1}^{n} \left \| \bar{x}_i - x_i \right \|^p \right )^{\frac{1}{p}}$.
For $p\mbox{--}norm$, studying the $\ell_0$, $\ell_2$ and $\ell_\infty$ is very common~\cite{Carlini017}. The $\ell_0$ is the count of changes in the adversarial example compared to the original sample, the $\ell_2$ is the Euclidean distance between the adversarial sample and the original sample, and $\ell_\infty$ is the maximum change with respect to all terms in the adversarial code examples.
In this work, we explore the size of perturbation with respect to the number of code lines and show the magnitude of such perturbation by $p\mbox{--}norm$ when considering non-targeted attacks.

\begin{figure}[t]
\centering
\begin{subfigure}[Small Code Perturbation \label{fig:norm1}]{\includegraphics[width=0.235\textwidth]{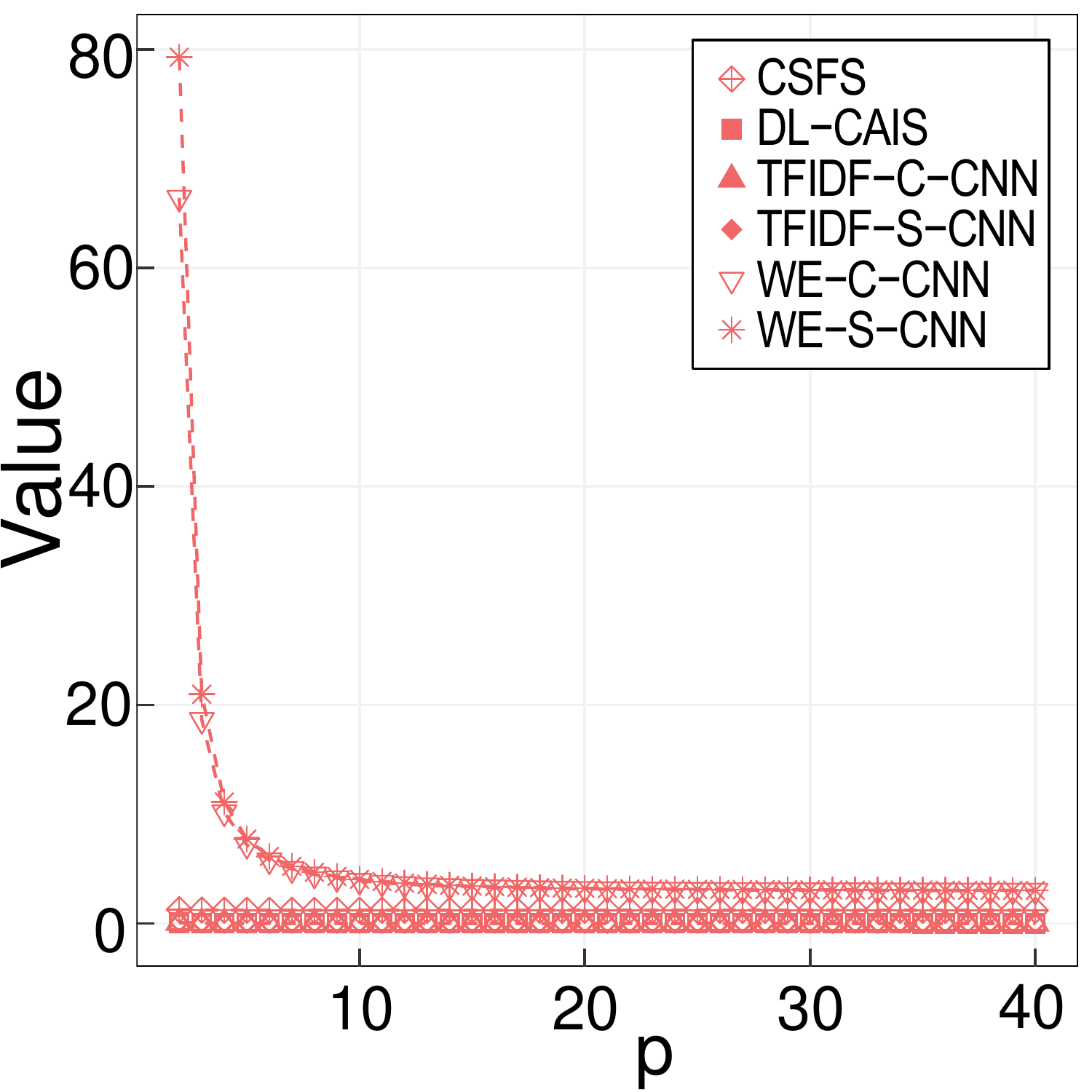}}
\end{subfigure}\vspace{-1mm}
\begin{subfigure}[Large Code Perturbation \label{fig:norm10}]{\includegraphics[width=0.235\textwidth]{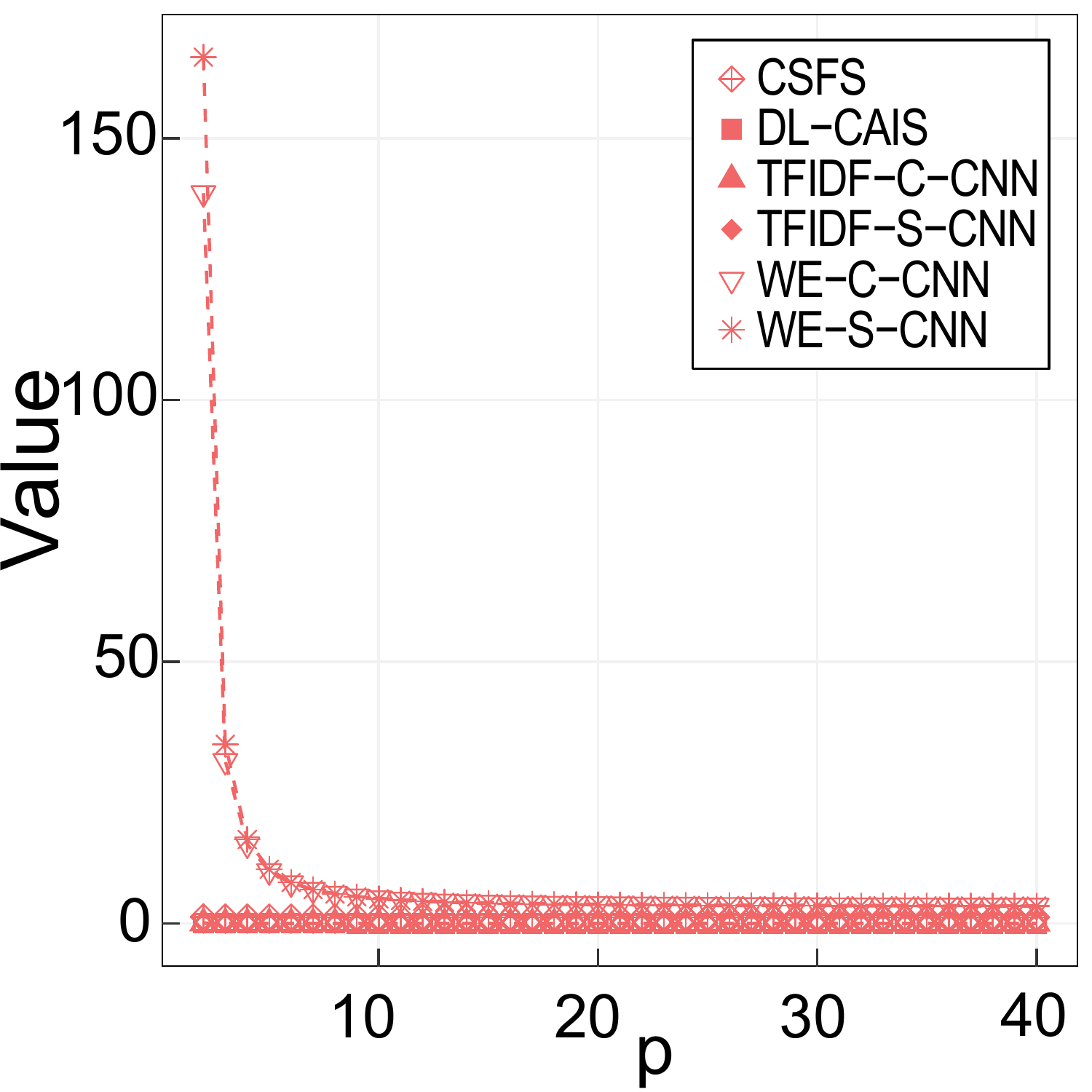}}
\end{subfigure}\vspace{-1mm}

\caption{The {\em p_norms} of perturbations with different sizes with respect to the initial representations of the original code. Small perturbations are generated by one statement, while large perturbations are generated by ten.} 
\label{fig:norms}\vspace{-5mm}
\end{figure}

Figure~\ref{fig:norms} shows the values of $p\mbox{--}norm$, ranging from $p=2$ to $p=40$, for small code perturbations generated from one statement of code and for large perturbations from ten code statements. The results shown in the figure are drawn from the average perturbation size using random perturbations generation (i.e.,  as used in the non-targeted attack) on a dataset of 100 programmers with nine files each.
Generally, the effect of perturbations on the input representations varies based on the underlying method used for representing the code. The effects are highest when using word embeddings as initial representations, e.g., WE-C-CNN and WE-S-CNN, due to the high-dimensional and compact representations of code sample using word embeddings, while the lowest effects are exhibited by approaches using TF-IDF as initial representations, e.g., DL-CAIS, TFIDF-C-CNN, and TFIDF-S-CNN.

Figure~\ref{fig:norm1} shows the $\ell_p$ of perturbations with one statement that includes only one line of code.
The $\ell_2$ values are:
0.07, 1.27, 0.08, 0.08, 66.43, 79.30 for
DL-CAIS, CSFS, WE-C-CNN, WE-S-CNN, TFIDF-C-CNN, and TFIDF-S-CNN, respectively. Increasing the value of $p$ decreases slightly the value of the $\ell_p$ to reach $\ell_{40} \approx 3$ for WE-C-CNN, and WE-S-CNN.
A similar observation is made using large perturbations regarding the effect on code representations. The $\ell_p$ values are shown to be slightly higher, which is due to the larger number of added code lines. Although these changes are not significant, they impact the outcome of models as seen in Figure~\ref{fig:untargeted_attack}.

\section{Related Work} \label{sec:related_work}
\BfPara{Code Authorship Attribution} Authorship attribution is a relatively rich topic, with a large number of proposed systems. The earlier work in this space attempted to define a set of stylistic traits and features that characterize authors of program~\cite{Krsul1997,843963,Brian2000}. Such traits may include byte- or gram-level attributes~\cite{Frantzeskou2006,abuhamad2021large}, control and data flow features~\cite{meng2017,rosenblum2011,alrabaee2014oba2},  and AST features~\cite{Brian2000,Caliskan-Islam:2015}.

Krsul and Spafford \cite{Krsul1997} proposed 
60 authorship features for C programmers, including the layout, programming style, and programming structure features, and posted 73\% of accuracy for identifying 29 programmers.  MacDonell \etal~\cite{843963} proposed 26 similar stylistic features for C++ authorship attribution and achieved 88\% of accuracy in identifying seven programmers. Frantzeskou \etal~\cite{Frantzeskou2006} introduced the concept of Author Profiles based on byte-level $n$-grams, allowing for 100\% accuracy in identifying eight programmers. 
Using a combination of $n$-grams and handpicked features, Burrows \etal~\cite{Burrows2009} scaled up authorship attribution to 100 programmers with 80.37\% of accuracy. The first large-scale study was introduced by Caliskan-Islam \etal~\cite{Caliskan-Islam:2015} using code stylometry features which allowed identifying 1,600 C++ programmers with 92.83\% accuracy. Abuhamad \etal \cite{abuhamad2018large} proposed DL-CAIS, which uses RNN-based models for deep representations, to identify 8,903 programmers with an accuracy of 92.3\%. Abuhamad \etal~\cite{abuhamad2019code} proposed a CNN-based approach that achieved an accuracy of 96.2\% for 1,600 programmers.
We consider the systems in~\cite{Caliskan-Islam:2015,abuhamad2018large,abuhamad2019code} with our attacks, and further details of their operation are in Section~\ref{sec:identification_workflow}.

\BfPara{Adversarial Authorship Attribution} Brennan \etal~\cite{brennan2012adversarial} proposed adversarial stylometry to circumvent authorship attribution of textual documents. The authors present a framework to create adversarial documents for two purposes: {\em obfuscation}, where an author of a document attempts to hide her identity, and {\em imitation}, where an author attempts to imitate the style of another author. Both approaches are conducted manually with human involvement.
For code authorship attribution,
Simko \etal~\cite{simko2018recognizing} pursued quantitative and qualitative approaches to evaluate authorship attribution under code forgeries. They recruited programmers to create code forgeries and human code analysts to evaluate and detect the forgeries. 
Meng \etal~\cite{meng2018adversarial} introduced attacks on binary-based authorship attribution using adversarial binaries that correspond to feature vector modifications calculated to meet various attack objectives. Matyukhina \etal~\cite{matyukhina2019adversarial} and Quiring \etal~\cite{quiring2019misleading} proposed a transformation process to hide or imitate programmers' coding style.  
Liu \etal~\cite{liu2021practical} proposed SCAD that uses a trained substitute model for the target identification classifier to evaluate a set of code transformations before probing the target model. These code transformations are generated based on a set of 37 transformation rules to change the code without changing its functionality. The transformations are then evaluated using a customized Jacobian-based saliency map approach based on the substitute classifier before querying the target model. All these methods require an exhaustive code transformations.
This aims to generate adversarial code examples using perturbations that serve the adversary's objectives without requiring code transformation.   
Another line of work, investigates the possible defenses against such attacks. Using adversarial training, Li \etal~\cite{li2022ropgen} proposed RoPGen, a method that combines data augmentation and gradient augmentation to enhance the diversity of training examples and learn distinct code attributions. RoPGen was able to improve the robustness of models against coding style imitation and hiding attacks.

\section{Conclusion} \label{sec:conclusion}

This work investigated the robustness of several code authorship attribution systems under different attacks utilizing code-level perturbation. 
We targeted six authorship attribution systems with different underlying techniques and defined different attack objectives---targeted and non-targeted. 
The attacks exploited code perturbations to hinder authorship recognition while preserving code functionality.
The process of generating AEs included producing code perturbations (targeted or non-targeted) to fulfill the adversaries' objectives.
Our results showed the impact of code perturbations on authorship attributions and how increasing the size of the perturbation increases its effect. However, the targeted techniques could be fooled with the smallest perturbations. All targeted systems were compromised under adversarial scenarios such that the confidence levels of the identification models were very low compared to the baseline performance.




\end{document}